# High Precision Renormalization Group Study of the Roughening Transition

**Martin Hasenbusch**[1], **Mihai Marcu**[2], **and Klaus Pinn**[3]

[1] CERN, Theory Division,
CH-1211 Genève 23, Switzerland

[2] Racah Institute of Physics,
Hebrew University, 91904 Jerusalem, Israel

[3] Institut für Theoretische Physik I, Universität Münster,
Wilhelm-Klemm-Str. 9, D-48149 Münster, Germany

**Abstract**

We confirm the Kosterlitz-Thouless scenario of the roughening transition for three different Solid-On-Solid models: the Discrete Gaussian model, the Absolute-Value-Solid-On-Solid model and the dual transform of the XY model with standard (cosine) action. The method is based on a matching of the renormalization group flow of the candidate models with the flow of a bona fide KT model, the exactly solvable BCSOS model. The Monte Carlo simulations are performed using efficient cluster algorithms. We obtain high precision estimates for the critical couplings and other non-universal quantities. For the XY model with cosine action our critical coupling estimate is $\beta_R^{XY} = 1.1197(5)$. For the roughening coupling of the Discrete Gaussian and the Absolute-Value-Solid-On-Solid model we find $K_R^{DG} = 0.6645(6)$ and $K_R^{ASOS} = 0.8061(3)$, respectively.

CERN-TH.7182/94
HU-RI-2/94
MS-TPI-94-3
April 1994

# Contents

# 1 Introduction

In 1951, Burton, Cabrera and Frank pointed out that a phase transition may occur in the equilibrium structure of crystal surfaces [1]. Such a phase transition from a smooth to a rough surface is called a *roughening transition*.

The roughening transition was observed in experiments by Heyraud and Métois on metal crystallites with a diameter of only a few micrometres [2]. More recent results for the roughening transition of a (110) surface of Ag were reported by Robinson et al. [3]. They found a roughening temperature $T_R = 790 \pm 20$ K. Several other groups have carried out beautiful experiments on roughening transitions for helium crystals in equilibrium with superfluid helium [4]. In these systems roughening temperatures of 0.35 K, 0.9 K and 1.28 K can be observed for the various inequivalent types of facets. Burton, Cabrera and Frank suggested that a growing layer of the crystal could approximately be described by the exactly soluble 2-dimensional Ising model. The idea was to represent the completed part of the layer by spin $+1$, while the vacancies are represented by spin $-1$. However, the model is quite unrealistic because it forbids the formation of a new layer before the previous one is completed.

A more appropriate description of a crystal in equilibrium with its vapor is given by the 3-dimensional Ising model, where a spin $+1$ represents a site occupied by an atom, while spin $-1$ represents a vacancy. The boundary conditions are chosen such that an interface forms between a region where a majority of spins are $+1$ and a region with most spins equal to $-1$. This interface is the model of the crystal surface. In 1973, Weeks et al. performed a low-temperature expansion for the width of an (100) interface in a 3-dimensional Ising model on a simple cubic lattice with isotropic couplings [5]. They found a roughening temperature $T_R = 0.57 \, T_c$. Here, $T_c$ denotes the critical temperature of the bulk phase transition. The approximation of the interface by the 2-dimensional Ising model yields $T_R = 0.503 \, T_c$.

A duality transformation exactly relates the 3-dimensional Ising model with the 3-dimensional $Z_2$ gauge theory. The roughening transition therefore also occurs in this gauge model [6, 7]. It was shown that also 4-dimensional lattice gauge models with continuous gauge group undergo a roughening transition [8–13]. The roughening transition is related to a restoration of symmetries broken by the introduction of the lattice.

A fairly good approximation of the Ising interface is given by Solid-On-Solid (SOS) models. The idea is to neglect overhangs of the interface and bubbles in the bulk. The variables of the 2-dimensional SOS models represent the height of the interface with respect to some reference plane.

SOS models are related with 2-dimensional XY models by a duality transformation [14]. The roughening transition of an SOS model corresponds to a transition of an XY model from a massive to a massless (spin wave) phase. In both representations, the transition is believed to be of the Kosterlitz-Thouless (KT) type [15]. The KT transition is a phase transition of infinite order, characterized by a very weak singularity in the free energy and an exponential singularity of the correlation length at the critical temperature. For reviews on the roughening transition see, e.g., [16].

Unfortunately for nearly all of the SOS models there is no rigorous proof that their phase transition is really of the KT type. For rigorous work on the existence of a phase transition from a massive to a massless phase, see [17]. See also [18], where the KT nature of the transition is put into question.

In order to confirm or reject the KT nature of the roughening transition, many Monte Carlo simulations [19] were performed of various SOS models and of the interface in the 3-dimensional Ising model [20–31].

Many Monte Carlo simulations were done to investigate the phase transition for the XY model with cosine or Villain action [32-49].

Most of the Monte Carlo studies of KT candidate models are based on a direct computation of critical quantities such as the correlation length or the susceptibility.

By fitting the data with different ansätze one tries to rule out the power law singularities of conventional phase transitions. However, it turns out to be very demanding to get data for sufficiently large correlation lengths with good statistics. Only the results of the latest simulations using cluster algorithms really favour a KT transition against a second-order phase transition, while the estimate for the transition temperature still has a relative error of order 1% [50, 51].

There have also been attempts to study the KT scenario with the help of the Monte Carlo Renormalization Group (MCRG). See, e.g., [52].

In this paper, we attack the problem with a new method.[1] This is based on the fact that one of the SOS models, the BCSOS model, can be solved exactly [55–57]. The BCSOS model has been *proved* to exhibit a KT transition. The critical coupling is exactly known. In addition, the correlation length and other quantities can be computed exactly [55].

---

[1] Part of this work is contained in [53]. A short report of our method and results is published in [54].

It was proposed long ago to improve the numerical study of SOS models by a comparison with BCSOS results [22]. In this report we give this comparison a precise meaning in the framework of the renormalization group (RG). We verify the KT scenario for several models - the ASOS model, the Discrete Gaussian (DG) model and the dual transform of the XY model with cosine action - by demonstrating that their long-distance RG flow at the critical point precisely matches with the flow of the critical BCSOS model. Stated differently, we demonstrate that the candidate models are in the same universality class (in the sense of Wilson's RG) as the BCSOS model.

The matching is demonstrated by comparing Monte Carlo data for expectation values of "blocked correlation functions". All data are generated using cluster Monte Carlo algorithms that do not suffer from CSD or have strongly reduced CSD. References for cluster algorithms for spin models are [58] – [61]. [58–61]. Cluster algorithms for SOS models were introduced and studied in [62–64]. Our RG comparison is designed in such a way that finite-size effects are exactly cancelled. By simulations on reasonably small lattices we obtain results for the critical couplings, which are competitive in precision with estimates from much more expensive Monte Carlo studies. We also get estimates for the non-universal constants determining the asymptotic behaviour of the correlation length.

The perhaps most important result of our study is, however, the demonstration that the models (with a high level of confidence) are in the same universality class as the BCSOS model. We consider this as an unambiguous confirmation that their phase transition is of the KT type.

This article is organized as follows: In section 2 we introduce the SOS models and their most important properties. In section 3 we summarize the heuristic picture of the KT phase transition using the flow diagram of Kosterlitz and Thouless. In section 4 we define an RG transformation for theories on finite lattices. The RG flow defined in section 4 lies at the heart of the matching analysis presented in section 5. There, we demonstrate that all the considered SOS models share the same universality class with the BCSOS model. Appendix 1 is devoted to the solution of the KT equations. In Appendix 2 we present the exact computation of the finite-size RG for the Gaussian model.

## 2   Solid-on-solid models

Solid-on-solid (SOS) models are of interest both theoretically and in the study of crystal interfaces. In this section we define some of these models and give an account of their most important properties. For reviews on SOS type of models we refer to [16].

All SOS models have in common that they are 2-dimensional lattice spin models. The spins $h_x$ take values in an unbounded discrete set $S$ (isomorphic to the integer numbers).

The interaction energy (Hamiltonian) is invariant under global shifts $h_x \to h_x + M$ for $M \in S$. A typical partition function for such a model looks like [2]

$$Z = \sum_{\{h\}} \exp\left(-\sum_{<x,y>} V(h_x - h_y)\right). \tag{1}$$

In this example, the Hamiltonian is a sum of contributions depending on pairs of nearest-neighbour spins only. The summation in eq. (1) is over equivalence classes of spin configurations $\{h\}$. The classes are defined by identifying two configurations that differ only by a global shift $M \in S$.

Our first example of an SOS model is the DG model: It is of the type defined in eq. (1) with

$$V_{DG} = K^{DG}(h_x - h_y)^2. \tag{2}$$

The spin variables $h_x$ take integer values. Note that the Hamiltonian looks exactly like that of a continuous Gaussian model. However, the restriction of the $h_x$ to integer values introduces a non-trivial interaction. Let us interpret the $h_x$ as heights with respect to a certain base. For finite positive $K^{DG}$ the Hamiltonian will favour that neighbouring spins take similar values. When $K^{DG}$ is large enough, the surface will not fluctuate too wildly. As a consequence one expects that the surface thickness squared,

$$\sigma^2 = \lim_{|x-y|\to\infty} \langle(h_x - h_y)^2\rangle, \tag{3}$$

is finite: the system is in the "smooth" phase. On the other hand, if $K^{DG}$ is below a certain critical value, the surface becomes "rough", and the surface thickness diverges.

In [17] it was proved for a class of SOS models that the 2-point correlation function $\langle(h_0 - h_x)^2\rangle$, for sufficiently small coupling $K$, goes like $\ln(|x|)$ at large distance $x$.

Furthermore, it follows from convergent cluster expansions that $\langle(h_0 - h_x)^2\rangle$ stays bounded for all $x$ if $K$ is sufficiently large.

On finite lattices with $N = L \times L$ sites we define

$$\sigma^2 = \frac{1}{N^2}\sum_{x,y}\langle(h_x - h_y)^2\rangle. \tag{4}$$

The surface thickness defined through this equation behaves exactly like the surface thickness defined via eq. (3): It stays finite for $L \to \infty$ if $K^{DG}$ is above a critical value, and it diverges with increasing $L$ if $K^{DG}$ is below this critical value. In this work, we shall refer to definition (4) throughout. The transition between the two phases is called roughening transition. The theory of Kosterlitz and Thouless makes detailed predictions about the nature of this transition, see section 3.

---

[2] A factor $1/k_B T$, where $k_B$ denotes Boltzmann's constant and $T$ the temperature, is absorbed in the definition of the interaction.

The Discrete Gaussian model is dual to the XY model with Villain action [14]. This model is defined by the partition function

$$Z_V = \int_{-\pi}^{\pi} \prod_x d\Theta_x \prod_{<x,y>} B(\Theta_x - \Theta_y), \tag{5}$$

with

$$B(\Theta) = \sum_{p=-\infty}^{\infty} \exp\left(-\tfrac{1}{2}\beta_V(\Theta - 2\pi p)^2\right) \tag{6}$$

and

$$\frac{1}{2\beta_V} = K^{DG}. \tag{7}$$

The index "$V$" here refers to "Villain".

The XY model with "standard (cosine) action" has the partition function

$$Z_{XY} = \int_{-\pi}^{\pi} \prod_x d\Theta_x \exp\left(\beta^{XY} \sum_{<x,y>} \cos(\Theta_x - \Theta_y)\right). \tag{8}$$

The standard action is the action discussed mostly for an XY model. The dual of this model is given by the partition function

$$Z_{XY}^{SOS} = \sum_{\{h\}} \prod_{<x,y>} I_{|h_x - h_y|}(\beta^{XY}), \tag{9}$$

where the $I_n$ are modified Bessel functions. Again $h_x$ is integer.

The Absolute-Value-Solid-On-Solid (ASOS) model is the SOS approximation of an interface in an Ising model on a simple cubic lattice on a (001)-lattice plane. It is defined by

$$V_{ASOS} = K^{ASOS} |h_x - h_y|. \tag{10}$$

We shall finally introduce the BCSOS (Body Centered Solid-On-Solid) model. It will play a prominent role in this paper. The BCSOS model was introduced by van Beijeren [65] as an SOS approximation of an interface in an Ising model on a body-centered cubic lattice on a (001)-lattice plane. The effective 2-dimensional lattice splits in two sublattices like a checker board. In the original formulation, on one of the sublattices the spins take integer values, whereas the spins on the other sublattice take half-integer values. We adopt a different convention: spins on "odd" lattice sites take values of the form $2n + \frac{1}{2}$, and spins on "even" sites are of the form $2n - \frac{1}{2}$, $n$ integer. As a consequence, the effective distribution for block spins (defined as averages over square blocks of size $B \times B$, where $B$ is an integer) will be centered around integer values instead of half-integer values, and the configurations with minimal energy will have integer average values, as is the case for

| $\kappa$ | $\xi_{\text{asympt}}$ | $\xi_{\text{exact}}$ | $\kappa$ | $\xi_{\text{asympt}}$ | $\xi_{\text{exact}}$ |
|---|---|---|---|---|---|
| 1.00 | 1.577 | 2.033 | 0.20 | $0.2436 \cdot 10^2$ | $0.2710 \cdot 10^2$ |
| 0.80 | 2.087 | 2.603 | 0.10 | $0.1750 \cdot 10^3$ | $0.1889 \cdot 10^3$ |
| 0.60 | 3.103 | 3.740 | 0.05 | $0.2784 \cdot 10^4$ | $0.2938 \cdot 10^4$ |
| 0.40 | 5.910 | 6.870 | 0.01 | $0.3080 \cdot 10^9$ | $0.3156 \cdot 10^9$ |

Table 1: Comparison of $\xi^{BCSOS}$ from the asymptotic and the exact formula

the other SOS models defined in this section. The partition function of the BCSOS model can be expressed as

$$Z_{BCSOS} = \sum_{\{h\}} \exp\left(-K^{BCSOS} \sum_{[x,y]} |h_x - h_y|\right) . \qquad (11)$$

The sum is over next-to-nearest-neighbour pairs $[x,y]$, and nearest-neighbour spins $h_x$ and $h_y$ obey the constraint $|h_x - h_y| = 1$. In [65] Van Beijeren showed that the BCSOS model is isomorphic to the F-model, which is a special six-vertex model. The configurations of the BCSOS model are in one-to-one correspondence to the configurations of the F-model. The F-model can be solved exactly with transfer matrix methods [56, 57, 55]. The roughening transition occurs at

$$K_R^{BCSOS} = \tfrac{1}{2} \ln 2 . \qquad (12)$$

The exact formula for the correlation length is [55]:

$$\begin{aligned} \frac{1}{\xi^{BCSOS}} &= -\ln\left\{2x^{1/2} \prod_{m=1}^{\infty} \left(\frac{1+x^{4m}}{1+x^{4m-2}}\right)^2\right\} , \\ x &\equiv \exp\left(-\text{arcosh}\left(\tfrac{1}{2}\exp(4K) - 1\right)\right) . \end{aligned} \qquad (13)$$

For $K \searrow K_R$, the correlation length behaves like

$$\xi^{BCSOS} \simeq \frac{1}{4} \exp\left(\frac{\pi^2}{8\sqrt{\tfrac{1}{2}\ln 2}} \kappa^{-\tfrac{1}{2}}\right) , \quad \kappa = \tfrac{K-K_R}{K_R} . \qquad (14)$$

The essential singularity of the correlation length at the critical coupling $K_R$ is typical for a KT model. It is instructive to compare the results of the exact and the asymptotic expression in the neighbourhood of the critical point. Table 1 shows that at correlation length 189 there is still a relative deviation of 7 % between the exact result and the result from the asymptotic formula. This nicely explains why the determination of amplitudes (occurring as parameters in the asymptotic correlation length formula) from correlation length measurements in KT models is so difficult.

The free energy per volume is also exactly known. For $K > K_R$,

$$f^{BCSOS} \equiv -\ln Z^{BCSOS}/\text{volume}$$

$$\begin{aligned} &= 2K - \left\{ \tfrac{1}{2}\lambda + \sum_{m=1}^{\infty} \frac{\exp(-m\lambda)\sinh(m\lambda)}{m\cosh(m\lambda)} \right\}, \\ \lambda &\equiv \operatorname{arcosh}\left(\tfrac{1}{2}\exp(4K) - 1\right). \end{aligned} \quad (15)$$

For $K < K_R$, the free energy $f^{BCSOS}$ has the following integral representation:

$$f^{BCSOS} = 2K - \frac{1}{4\mu}\int_0^\infty \frac{dx}{\cosh(\pi x/2\mu)} \ln\left(\frac{\cosh x - \cos 2\mu}{\cosh x - 1}\right), \quad (16)$$

$$\mu \equiv \arccos\left(\tfrac{1}{2}\exp(4K) - 1\right). \quad (17)$$

Figure 1 shows $f^{BCSOS}$ in the neighbourhood of the roughening coupling $K_R = \tfrac{1}{2}\ln 2$. The free energy and all its derivatives stay finite at the transition point (when suitably defined as limits from the right and from the left, respectively). However, there is an essential singularity: This can already be seen if one directly inserts $K = K_R$ in eqs. (15) and (16). In both cases one obtains $\ln 2$, to be compared with the right-left limit in fig. 1. It can be shown [55] that the singular part of the free energy in the vicinity of the roughening point is of the form

$$f_{\text{sing}}^{BCSOS} \sim \exp\left(-\frac{\pi^2}{4\sqrt{\tfrac{1}{2}\ln 2}} \kappa^{-\tfrac{1}{2}}\right), \quad \kappa = \frac{K - K_R}{K_R}. \quad (18)$$

## 3 Kosterlitz-Thouless theory

The Kosterlitz-Thouless theory is conveniently discussed in the framework of the Sine Gordon model, a system with continuous spins and lattice Hamiltonian

$$\mathcal{H}(\varphi) = \frac{1}{2\beta} \sum_{<x,y>} (\phi_y - \phi_y)^2 - z \sum_x \cos(2\pi\varphi_x). \quad (19)$$

The coupling constant $z$ is called "fugacity", because the model can be exactly transformed into a 2-dimensional Coulomb gas, where $z$ plays the role of the fugacity for charged particles. In the limit of vanishing fugacity the roughening coupling of the Sine Gordon model is

$$\lim_{z \to 0} \beta_R^{SG}(z) = \frac{2}{\pi}. \quad (20)$$

The critical behaviour of the model is determined by the long-distance properties of the correlation functions. The most systematic way to analyze long-distance properties of a statistical model is the renormalization group as pioneered by Kadanoff et al. [66] and worked out systematically by Wilson [67]. At the heart of the renormalization group approach to critical phenomena lies the notion of the effective Hamiltonian, which encodes the properties of the model at length scales larger than the (microscopic) scale of the fundamental Hamiltonian. For a lattice model the effective Hamiltonian $\mathcal{H}_{\text{eff}}$ is defined through[3]

$$\exp(-\mathcal{H}_{\text{eff}}(\phi)) = \int \prod_{x \in \Lambda} d\varphi_x \exp(-\mathcal{H}(\varphi)) \prod_{x' \in \Lambda'} \delta\left(\phi_{x'} - B^{-2} \sum_{x \in x'} \varphi_x\right). \quad (21)$$

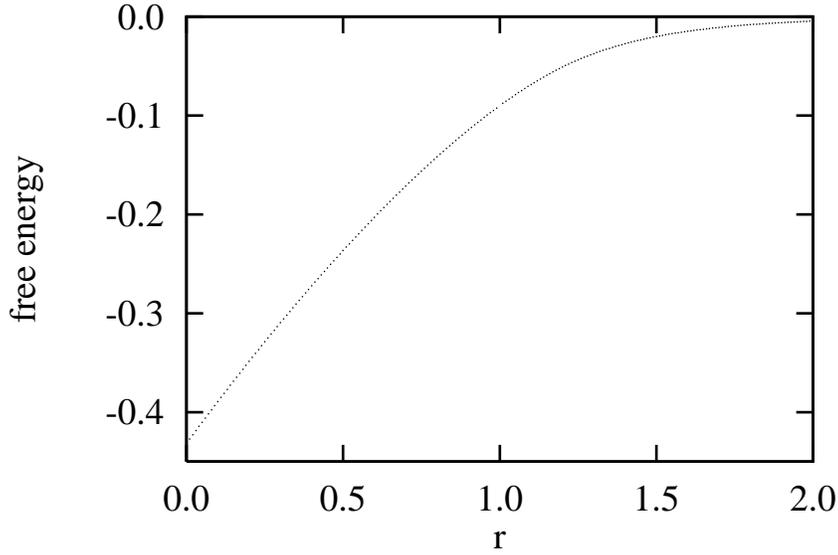

Figure 1: The free energy per volume of the BCSOS model as a function of $r = K^{BCSOS}/K_R^{BCSOS}$

The effective theory lives on a lattice $\Lambda'$, which has lattice spacing $B$ (in units of the original lattice spacing). The block spins $\phi_{x'}$ are defined as averages of the original spins $\varphi_x$ of square blocks $x'$ of size $B \times B$.

In general, one needs a non-trivial rescaling of the block spin field (in accordance with a non-vanishing critical exponent $\eta$). However, for the **Z**-symmetric models studied in this work, there is no anomalous dimension: Note that for SOS models the effective Hamiltonians defined through eq. (21) also have a global **Z**-symmetry. This property would be destroyed by introducing additional rescaling in the definition of block spin variables [4].

A renormalization group flow is defined by iterating the renormalization group transformation. In the case of a linear block spin definition the iteration of a single step is exactly equivalent to going to larger and larger block size. The property $R(B_1)\, R(B_2) = R(B_1\, B_2)$ is called *half group property*.

---
[3]There exist, of course, many other possible definitions of an effective Hamiltonian.

[4]The critical fixed point, which is a continuous Gaussian model, has even larger symmetry, namely the real numbers **R**. This phenomenon is called *symmetry enhancement*.

One of the practical difficulties one has to deal with is the proliferation of coupling constants: the exact effective Hamiltonian can no longer be exactly parametrized by a finite number of coupling constants. There is an infinite number of effective coupling constants. However, one assumes that only a few couplings are important for the critical behaviour. This has of course to be demonstrated. The study of the effects from truncating the effective Hamiltonian is a difficult problem.

KT theory is based on the assumption that the RG flow of the critical or nearly critical Sine Gordon model (and also of other KT models) at long distance is well described by two parameters $\beta$ and $z$ [15].

The study of the flow of $\beta$ and $z$ with increasing length scale is most convenient in a continuum formulation, where the lattice cutoff is replaced by a momentum space cutoff. A length scale is then identified with the inverse momentum cutoff. The cutoff can be changed continuously. One can derive differential equations for the flow of the effective Hamiltonian [15]. The KT flow equations are

$$\begin{aligned} \dot{y} &= -xy\,, \\ \dot{x} &= -y^2\,, \end{aligned} \qquad (22)$$

where the dot denotes the derivative with respect to the scale parameter $t$. The corresponding length scale is $B = \exp(t)$, and $x$ and $y$ are related to $\beta$ and $z$ by

$$\begin{aligned} x &= \pi\beta - 2\,, \\ y &= c\,z\,, \end{aligned} \qquad (23)$$

where the constant $c$ is non-universal, i.e. depends on the choice of the regularization procedure. The equations can be exactly solved, see Appendix 1. The solutions satisfy the equation $y^2 = E + x^2$, or

$$y = \sqrt{E + x^2}\,. \qquad (24)$$

Typical renormalization group trajectories are shown in fig. 2. The flow diagram encodes the full information about the critical behaviour. Region I ($E < 0$) corresponds to the massless phase. Here, the trajectories run exponentially fast (in the variable $t = \ln B$) to $y = 0$: the long-distance behaviour is that of a massless Gaussian model. The point where a trajectory hits the $x$-axis corresponds to a $\beta$-value that is called $\beta_{\text{eff}}$. It is the $\beta$-coupling of the free field theory describing the long distance behaviour of all models that "come down" the trajectory. Note that this is a whole class of models, which differ only in the "time" they need to reach the $x$-axis.

Regions II ($E > 0$) and III ($E < 0$) correspond to the massive phase. Here, with increasing length scale, the fugacity increases, and the global symmetry under shifts $\varphi_x \to \varphi_x +$integer is spontaneously broken. The separatrix between regions I and II is the critical line. Models on this line are driven into the fixed point $(x, y) = (0, 0)$, which corresponds to $\beta = 2/\pi$ and $z = 0$. Note that the approach to the fixed point goes like $\sim t^{-1}$ on the critical trajectory.

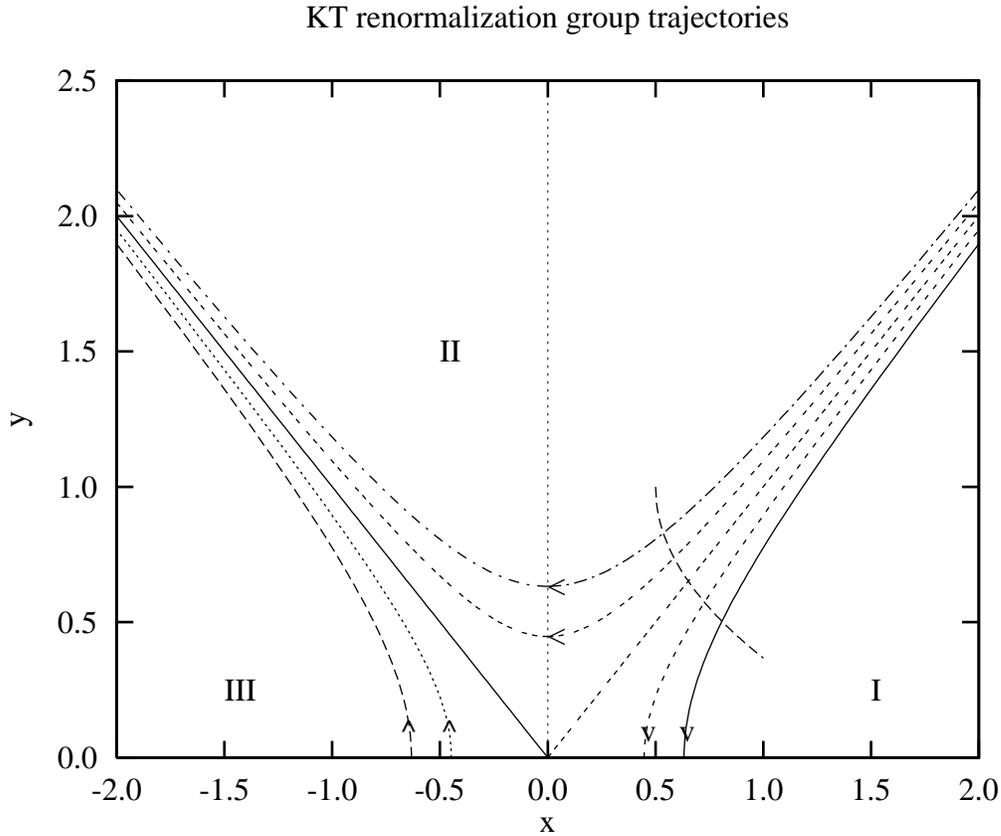

Figure 2: Renormalization group trajectories in a Kosterlitz-Thouless model. Region I ($E < 0$) corresponds to the massless phase. Here, the trajectories run exponentially fast (in the variable $t$) to $y = 0$: the long-distance behaviour is that of a Gaussian model. Regions II ($E > 0$) and III ($E < 0$) correspond to the massive phase. Here, with increasing length scale, the fugacity increases, and the global symmetry under shifts $\varphi_x \to \varphi_x +$ integer is spontaneously broken. The separatrix between regions I and II ($E = 0$) is the critical line. Models on this line are driven into the fixed point $(x, y) = (0, 0)$ which corresponds to $\beta = 2/\pi$ and $z = 0$. The meaning of the dotted line intersecting the trajectories in regions I and II is explained in the text.

From the flow diagram one can derive (see, e.g., [68]) that the correlation length of a KT model diverges like

$$\xi^{KT} \simeq A \, \exp\left( C \, \left|\frac{\beta_c - \beta}{\beta_c}\right|^{-\frac{1}{2}} \right) . \qquad (25)$$

Here, $A$, $C$, and $\beta_c$ are non-universal constants, i.e. they have different values for different KT models. Compare with eq. (14).

The KT equations for the Sine Gordon model are derived using several approximations; the most important of these is that the fugacity is small. However, if one assumes that $(x, y) = (0, 0)$ is the relevant fixed point for the critical behaviour, then the trajectories of a critical or nearly critical KT model will eventually enter the region where the fugacity is small and where the KT equations become valid. Depending on the details of the microscopic Hamiltonian, the "time" before the effective Hamiltonian comes close to the fixed point can be very long, i.e. the flow close to the critical line is very slow. This is one of the reasons why the unambiguous confirmation of the KT scenario for realistic models is so difficult.

Idealizing things very much, one of the SOS models introduced in section 2 can be represented by a line like the dotted line in fig. 2 intersecting the trajectories in regions I and II. The line should be carefully interpreted as follows: to each value of the SOS coupling $K$, there corresponds one of the trajectories in the flow diagram the model eventually will run on after several RG steps.

## 4 Finite lattice renormalization group

Universal properties of a statistical system do not depend on short distance details, but only on the nature of long wavelength fluctuations. This suggests to remove the irrelevant high frequency degrees of freedom by applying a coarse graining (block spin) procedure as introduced in section 3.

Universality, which was first introduced as the coincidence of the critical indices of various models, can be expressed as a convergence of the renormalization group flow to a universal flow as $T \to T_R$ and the number of block spin transformations goes to infinity. Here we use $T$ as a representative for any coupling that is driven to a critical value in order to make the correlation length diverge and the system become critical.

Consider two models with different microscopic Hamiltonians that belong to the same universality class, i.e. that have the same critical indices. The above statement says that if the two models are both at criticality then their effective Hamiltonians will converge towards the same fixed point Hamiltonian. It might even happen that the two flows of Hamiltonians will come close to each other already a long time before they are really close to the fixed point. (We shall actually observe this phenomenon for the models studied in this paper.) Note, however, that the two systems might need a different number of RG steps to reach a certain point on the universal trajectory.

It is suggestive to compare renormalization group flows in order to test for universality properties. To directly compare effective Hamiltonians, one would have to parametrize them in terms of coupling constants. However, there are infinitely many couplings already after a single renormalization group step. It turns out to be very difficult to determine the coupling constants of the blocked system by analytical calculations [69] or Monte Carlo simulations (MCRG) [70].

We shall deal with the problem in a similar fashion as Shenker and Tobochnik [71] did for the 2-dimensional O(3) model and Wilson [72] for the 4-dimensional SU(2) gauge model. We define effective Hamiltonians for *finite* lattices $\Lambda$ that consist of $l \times l$ blocks, where each of the blocks contains $B \times B$ sites. The effective Hamiltonian $\mathcal{H}_{\text{eff}}^{(l,B)}$ defined through

$$\exp(-\mathcal{H}_{\text{eff}}^{(l,B)}(\phi)) = \int \prod_{x \in \Lambda} d\varphi_x \exp(-\mathcal{H}(\varphi)) \prod_{x' \in \Lambda'} \delta \left( \phi_{x'} - B^{-2} \sum_{x \in x'} \varphi_x \right) \qquad (26)$$

is a function of $l^2$ block spin variables $\phi_{x'}$. A renormalization group flow is now defined as the sequence of effective Hamiltonians $\mathcal{H}_{\text{eff}}^{(l,B)}$, for fixed $l$ and increasing $B$. Note that different $l$'s lead to different flows.

In order to monitor the flow of the $\mathcal{H}_{\text{eff}}^{(l,B)}$, we do not necessarily have to compute the still infinitely many couplings in $\mathcal{H}_{\text{eff}}^{(l,B)}$. We can instead consider a set of suitably chosen observables,

$$E_i^{(l,B)} \equiv \langle A_i(\phi) \rangle_{l,B} = Z^{-1} \int D\phi \, \exp(-\mathcal{H}_{\text{eff}}^{(l,B)}) \, A_i(\phi) \,. \qquad (27)$$

A convergence of the flow of effective Hamiltonians $\mathcal{H}_{\text{eff}}^{(l,B)}$ towards a fixed point will imply also the convergence of the flows of the $E_i^{(l,B)}$. The crucial point is that the $E_i^{(l,B)}$ can be expressed as expectation values in the original system with Hamiltonian $\mathcal{H}$ on a lattice with size $L = lB$. One just has to measure correlation functions of block averages of the original system, that can be simulated, e.g. with efficient Monte Carlo algorithms.

It is worth noting that the couplings $E_i^{(l,B)}$ introduced above can be interpreted as *phenomenological couplings* as introduced by Nightingale and by Binder, see [73].

We conclude this section by presenting results for the flow of the $\mathcal{H}_{\text{eff}}^{(l,B)}$ for the free massless field theory (Gaussian model) in two dimensions. This may serve as an illustration for the convergence of the flow towards a fixed point. The Hamiltonian is

$$\mathcal{H}(\varphi) = \tfrac{1}{2}(\varphi, -\Delta\varphi) = \tfrac{1}{2} \sum_{<x,y>} (\varphi_x - \varphi_y)^2 \,. \qquad (28)$$

For this theory the effective Hamiltonian can be computed exactly (see Appendix 2),

$$\mathcal{H}_{\text{eff}}^{(l,B)}(\phi) = \frac{1}{2} \left( \phi, -\Delta_{\text{eff}}^{(l,B)} \phi \right) = \tfrac{1}{2} \sum_{x',y'} \phi_{x'} \left[ -\Delta_{\text{eff}}^{(l,B)} \right]_{x',y'} \phi_{y'} \,. \qquad (29)$$

| $B$ | $a$ | $b$ | $c$ | $d$ | $e$ | $f$ |
|---|---|---|---|---|---|---|
| 4 | $-8.99182$ | 2.98941 | $-0.12121$ | $-1.10822$ | $-0.08032$ | 0.05674 |
| 8 | $-9.83928$ | 3.41193 | $-0.22512$ | $-1.38204$ | $-0.05261$ | 0.06654 |
| 16 | $-10.0911$ | 3.54303 | $-0.26380$ | $-1.46882$ | $-0.03870$ | 0.06655 |
| 32 | $-10.1573$ | 3.57796 | $-0.27464$ | $-1.49208$ | $-0.03453$ | 0.06626 |
| 64 | $-10.1740$ | 3.58685 | $-0.27744$ | $-1.49801$ | $-0.03343$ | 0.06616 |
| 128 | $-10.1783$ | 3.58908 | $-0.27815$ | $-1.49950$ | $-0.03315$ | 0.06613 |
| 256 | $-10.1793$ | 3.58964 | $-0.27833$ | $-1.49988$ | $-0.03308$ | 0.06612 |
| 512 | $-10.1796$ | 3.58978 | $-0.27837$ | $-1.49997$ | $-0.03307$ | 0.06612 |
| 1024 | $-10.1796$ | 3.58982 | $-0.27838$ | $-1.49999$ | $-0.03306$ | 0.06612 |
| 2048 | $-10.1797$ | 3.58983 | $-0.27838$ | $-1.50000$ | $-0.03306$ | 0.06612 |

Table 2: Flow of the effective Laplacian for $l = 4$. The components are arranged as indicated in eq. (30)

Table 2 shows the components of $\Delta_{\text{eff}}^{(l,B)}$ on a $4 \times 4$ lattice, which are arranged according to the following scheme:

$$\begin{array}{cccc} a & b & d & b \\ b & c & e & c \\ d & e & f & e \\ b & c & e & c \end{array} \tag{30}$$

## 5  RG matching with the BCSOS model

Using the cluster algorithm described in [63], we simulated the BCSOS model at the roughening coupling $K_R^{BCSOS} = \frac{1}{2} \ln 2$ on square lattices of size $L \times L$ with periodic boundary conditions. For a list of the used $L$'s and the statistics, see table 3. There we give as an example the results for the squared interface width $\sigma^2$ defined in eq. (4).

| $L$ | $\sigma^2$ | $stat$ | $L$ | $\sigma^2$ | $stat$ |
|---|---|---|---|---|---|
| 12 | 1.02860(32) | 2.8 | 48 | 1.34674(34) | 2.2 |
| 16 | 1.09624(42) | 1.4 | 64 | 1.41080(26) | 4.4 |
| 24 | 1.18996(40) | 1.6 | 96 | 1.50038(30) | 5.3 |
| 32 | 1.25464(38) | 1.5 | 128 | 1.56362(32) | 4.9 |

Table 3: Results for the squared interface width $\sigma^2$ as function of the lattice size $L$ for the critical BCSOS model. The statistics $stat$ is given in units of $10^6$ single cluster updates

Figure 3 shows $\sigma^2 - (2/\pi^2) \ln L$ as a function of $\ln L$. KT theory predicts that this quantity should converge towards a constant for large $L$, see [15]. The figure shows a significant deviation from this behaviour. This means that for the lengths fitting on lattices with $L \leq 128$ the effective fugacity is not small. This is a consequence of the fact that along the critical line the flow towards the Gaussian fixed point is very slow (like $\sim 1/\ln L$, see section 3).

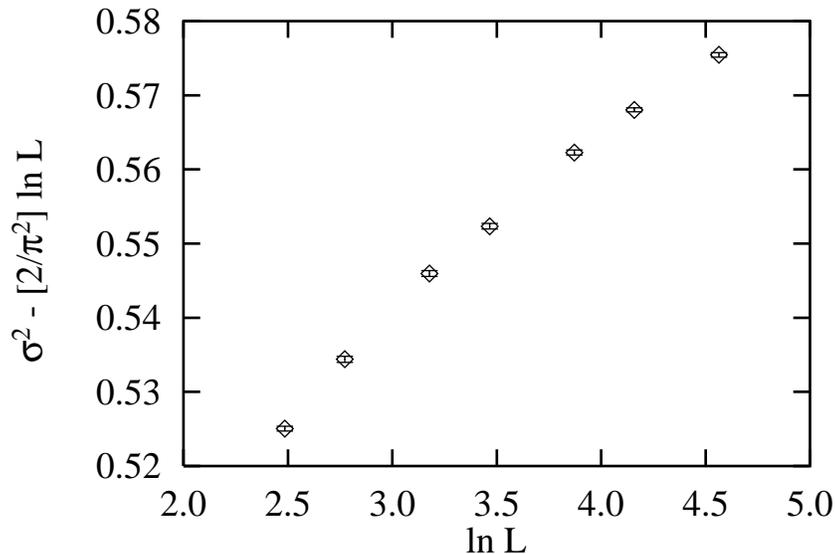

Figure 3: The squared surface width $\sigma^2$ of the BCSOS model minus its large $L$ behaviour as anticipated from KT theory

Our method to monitor the RG flow is to compute the flow of blocked observables (see section 4). The lattice is divided into $l \times l$ blocks of size $B \times B$, with $l = 1, 2, 4$.[5] Linear block spins $\phi_{x'}$ are defined according to eq. (21).

Motivated by KT theory we measured two types of block observables: those that are sensitive to the flow of the kinetic term (flow of $K$), and those that are sensitive to the fugacity. For the first type of observables we chose

$$A_{1,l} = \left\langle \frac{1}{2l^2} \sum_{<x',y'>} (\phi_{x'} - \phi_{y'})^2 \right\rangle , \qquad (31)$$

where $<x', y'>$ are nearest-neighbour pairs on the block lattice, and

$$A_{2,l} = \left\langle \frac{1}{2l^2} \sum_{[x',y']} (\phi_{x'} - \phi_{y'})^2 \right\rangle , \qquad (32)$$

---

[5] A posteriori we found that measuring also $l > 4$ data would have been useful. However, the small $l$-values enabled us to save all block spin configurations on disk. This gave us flexibility in the data analysis.

| $L$ | $A_{1,2}^{(0)}$ | $A_{2,2}^{(0)}$ | $A_{1,4}^{(0)}$ | $A_{2,4}^{(0)}$ |
|---|---|---|---|---|
| 8 | 0.136719 | 0.187500 | 0.293527 | 0.380581 |
| 12 | 0.126721 | 0.175926 | 0.257225 | 0.340481 |
| 16 | 0.123147 | 0.171875 | 0.243918 | 0.326090 |
| 24 | 0.120565 | 0.168981 | 0.234147 | 0.315663 |
| 32 | 0.119655 | 0.167969 | 0.230662 | 0.311978 |
| 48 | 0.119002 | 0.167245 | 0.228148 | 0.309333 |
| 64 | 0.118773 | 0.166992 | 0.227263 | 0.308404 |
| 96 | 0.118609 | 0.166811 | 0.226628 | 0.307739 |
| 128 | 0.118551 | 0.166748 | 0.226406 | 0.307507 |
| 256 | 0.118496 | 0.166687 | 0.226191 | 0.307282 |
| 512 | 0.118482 | 0.166672 | 0.226137 | 0.307226 |
| 1024 | 0.118479 | 0.166668 | 0.226124 | 0.307212 |
| 2048 | 0.118478 | 0.166667 | 0.226120 | 0.307208 |
| 4096 | 0.118478 | 0.166667 | 0.226119 | 0.307207 |
| 8192 | 0.118478 | 0.166667 | 0.226119 | 0.307207 |

Table 4: Exact results for $A_{1,l}^{(0)}$ and $A_{2,l}^{(0)}$

where $[x', y']$ are next-to-nearest-neighbour pairs. These quantities are defined for $l > 1$. For the actual matching procedure to be described below, we also employed the quantities

$$D_{i,l} = \frac{A_{i,l}^{(0)}|_{B=\infty}}{A_{i,l}^{(0)}} A_{i,l} \quad \text{for } l = 1, 2 \,. \tag{33}$$

Here, the $A_{i,l}^{(0)}$ denote the same quantities as defined in eqs. (31) and (32), taken, however, in the continuous Gaussian model with the Hamiltonian defined in eq. (28). The $A_{i,l}^{(0)}$ can be computed exactly, see Appendix 2. In table 4 we give the values of $A_{1,l}^{(0)}$ and $A_{2,l}^{(0)}$ for $L \leq 8192$. Within the accuracy obtained (6 digits), the infinite $B$ limit is reached for $L = 4096$.

As a monitor for the fugacity we chose the following set of quantities (defined for $l = 1, 2, 4$):

$$\begin{aligned}
A_{3,l} &= \left\langle \frac{1}{l^2} \sum_{x'} \cos(1 \cdot 2\pi \phi_{x'}) \right\rangle, \\
A_{4,l} &= \left\langle \frac{1}{l^2} \sum_{x'} \cos(2 \cdot 2\pi \phi_{x'}) \right\rangle, \\
A_{5,l} &= \left\langle \frac{1}{l^2} \sum_{x'} \cos(3 \cdot 2\pi \phi_{x'}) \right\rangle.
\end{aligned} \tag{34}$$

We believe that all important information about the large-distance RG flow of an SOS model close to or at criticality can be monitored by these observables or by a subset of them. The result for the flow of the $A's$ for the critical BCSOS model is summarized in table 5. We also give the exact limits that these quantities should approach when the block size $B$ is scaled to $\infty$. KT theory predicts that $A_{3,l}$, $A_{4,l}$ and $A_{5,l}$ have to converge to zero. The quantities $A_{1,l}$ and $A_{2,l}$ are predicted to converge to $2/\pi$ times the $B \to \infty$ limit of the same observables in the free field theory. These limits can be read off from table 4. A close look at the data reveals that even for large $B$ the $A's$ are still off their fixed points values. However, we shall see in the following that it does not matter that in the flow of the BCSOS data the fixed point is still somewhat away: The RG matching will take place a long time before the fixed point is close. Irrelevant couplings die out with a power of the length scale. The flow of the couplings is rapidly reduced to a 1-dimensional manifold. Along this remaining line, the fugacity dies out logarithmically with the length scale. Eventually the Gaussian fixed point with zero fugacity is reached. Since we know the 1-dimensional manifold from the BCSOS model, we just have to recover it in the other models. This can be done even far away from the fixed point.

| $L$ | $l$ | $B$ | $A_{1,l}$ | $A_{2,l}$ | $A_{3,l}$ | $A_{4,l}$ | $A_{5,l}$ |
|---|---|---|---|---|---|---|---|
| 12 | 1 | 12 | | | 0.2842(17) | 0.0777(14) | 0.0229(13) |
| 16 | 1 | 16 | | | 0.2655(27) | 0.0669(20) | 0.0212(18) |
| 24 | 1 | 24 | | | 0.2383(31) | 0.0551(22) | 0.0149(19) |
| 32 | 1 | 32 | | | 0.2261(36) | 0.0511(24) | 0.0105(20) |
| 48 | 1 | 48 | | | 0.2056(36) | 0.0394(23) | 0.0077(18) |
| 64 | 1 | 64 | | | 0.1969(30) | 0.0372(18) | 0.0051(13) |
| 96 | 1 | 96 | | | 0.1821(37) | 0.0296(20) | 0.0052(14) |
| 128 | 1 | 128 | | | 0.1719(43) | 0.0283(21) | 0.0058(15) |
| $\infty$ | 1 | $\infty$ | | | 0.0 | 0.0 | 0.0 |
| 12 | 2 | 6 | 0.08891(16) | 0.11983(26) | 0.2403(10) | 0.0668(7) | 0.0227(6) |
| 16 | 2 | 8 | 0.08569(22) | 0.11745(36) | 0.2258(16) | 0.0582(10) | 0.0157(9) |
| 24 | 2 | 12 | 0.08319(22) | 0.11505(37) | 0.2009(19) | 0.0445(11) | 0.0101(9) |
| 32 | 2 | 16 | 0.08187(23) | 0.11395(39) | 0.1892(22) | 0.0390(11) | 0.0084(10) |
| 48 | 2 | 24 | 0.08114(22) | 0.11343(37) | 0.1746(22) | 0.0328(11) | 0.0058(9) |
| 64 | 2 | 64 | 0.08097(17) | 0.11339(29) | 0.1655(18) | 0.0263(8) | 0.0042(7) |
| 96 | 2 | 48 | 0.08076(20) | 0.11326(33) | 0.1528(22) | 0.0239(9) | 0.0046(7) |
| 128 | 2 | 64 | 0.08038(21) | 0.11268(35) | 0.1435(26) | 0.0208(10) | 0.0033(7) |
| $\infty$ | 2 | $\infty$ | 0.075425 | 0.106104 | 0.0 | 0.0 | 0.0 |
| 12 | 4 | 3 | 0.19524(17) | 0.24472(24) | 0.2438(5) | 0.1121(4) | 0.0586(4) |
| 16 | 4 | 4 | 0.17686(21) | 0.23046(31) | 0.2226(7) | 0.0781(5) | 0.0407(5) |
| 24 | 4 | 6 | 0.16521(20) | 0.21978(30) | 0.1933(8) | 0.0535(6) | 0.0177(5) |
| 32 | 4 | 8 | 0.16087(21) | 0.21539(31) | 0.1793(9) | 0.0438(6) | 0.0122(5) |
| 48 | 4 | 12 | 0.15798(18) | 0.21278(28) | 0.1632(9) | 0.0349(5) | 0.0087(4) |
| 64 | 4 | 16 | 0.15646(14) | 0.21136(21) | 0.1522(8) | 0.0296(4) | 0.0064(3) |
| 96 | 4 | 24 | 0.15522(16) | 0.20995(25) | 0.1395(9) | 0.0248(5) | 0.0049(4) |
| 128 | 4 | 32 | 0.15471(17) | 0.20928(27) | 0.1322(11) | 0.0219(5) | 0.0045(4) |
| $\infty$ | 4 | $\infty$ | 0.143952 | 0.195574 | 0.0 | 0.0 | 0.0 |

Table 5: Finite lattice renormalization group flow of the $A_{i,l}$ for the critical BCSOS model

The simulations of the DG model, the ASOS model and the dual of the XY model were also performed on quadratic lattices with periodic boundary conditions. We used the very efficient VMR cluster algorithms [62]. The same blocking prescription as for the BCSOS model was employed. The lattice sizes and couplings involved will be specified below.

## 5.1 Determination of the roughening couplings

There are two parameters that have to be tuned in order to match the RG flow of one of the SOS models with that of the critical BCSOS model. One can vary the coupling $K^{SOS}$ of the SOS model. This allows one to walk on the approximate starting line in fig. 2. The flow of the SOS model can only match that of the critical BCSOS model if $K^{SOS} = K_R^{SOS}$. On the other hand, one can vary the ratio of the lattice sizes of the SOS model and the BCSOS model and, as a consequence, the ratio of the block sizes $b_m^{SOS} = B^{SOS}/B^{BCSOS}$. A ratio $b_m^{SOS} \neq 1$ turns out to be necessary to compensate for the different positions of the approximate starting lines in fig. 4.

Before we turn to the question of how to determine $K_R^{SOS}$ and $b_m^{SOS}$ in practice, let us write down the general condition for one of the SOS models to be in the same universality class as the critical BCSOS model.

<u>Matching condition:</u> Universality holds if there exists a $b_m$ and a $K_R^{SOS}$ such that for all $i, l$,

$$A_{i,l}^{SOS}\left(b_m^{SOS} B^{BCSOS}, K_R^{SOS}\right) = A_{i,l}^{BCSOS}\left(B^{BCSOS}, K_R^{BCSOS}\right) \tag{35}$$

in the limit of large $B^{BCSOS}$, and the corrections are of order $(B^{BCSOS})^{-\omega}$, with $\omega > 0$ the leading correction to scaling exponent. As we shall see below, moderate $B^{BCSOS}$ are sufficient in practice.

In order to determine the SOS roughening coupling $K_R^{SOS}$ from the RG flow data we proceeded as follows: For fixed values of $L^{SOS}$ and for fixed $l$, we considered the two equations:

$$\begin{aligned}
A_{1,l}^{SOS}\left(B^{SOS}, K_1^{SOS}\right) &\equiv A_{1,l}^{BCSOS}\left(B^{BCSOS}, K_R^{BCSOS}\right), \\
A_{3,l}^{SOS}\left(B^{SOS}, K_3^{SOS}\right) &\equiv A_{3,l}^{BCSOS}\left(B^{BCSOS}, K_R^{BCSOS}\right).
\end{aligned} \tag{36}$$

We chose $A_1$ and $A_3$ because we consider these to be the most important observables for the monitoring of the RG flow. For each of the available values of $B^{BCSOS}$ listed in table 5 we solved these two equations for the couplings $K_i^{SOS}$. To be able to do this we needed the $A$'s for a range of couplings. We simulated the SOS models at the (to that time) best known estimate for their roughening coupling. The expectation values in a neighbourhood of the simulation point could then be obtained by extrapolating using a reweighting method [74].

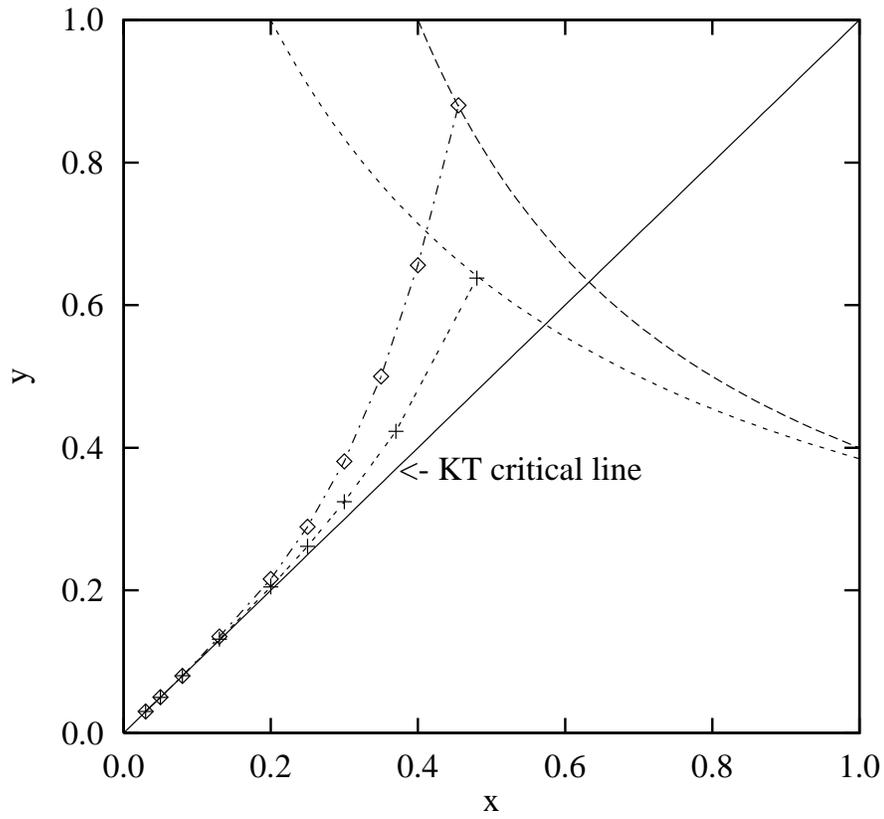

Figure 4: Matching of the finite size RG flow of two different SOS models in the KT flow diagram. Two successive points along the discrete trajectories are separated by a fixed scale factor. (In reality the models do not 'start' in the KT diagram but rather in a higher dimensional space of coupling constants.) Note that one of the models is one step 'ahead' of the other one. This explains that one needs the offset factor $b_m$ in order to match the flows.

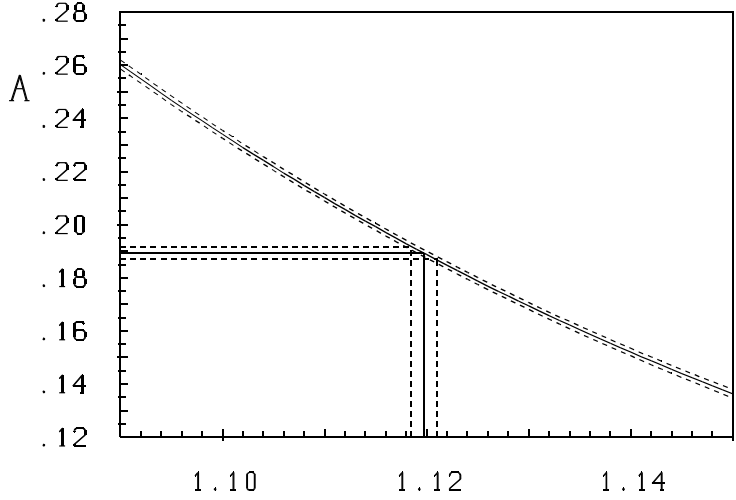

Figure 5: Determination of $\beta_3^{XY}$ for $L^{XY} = 32$, $L^{BCSOS} = 32$ and $l = 2$. The solid curve gives $A_{3,2}^{XY}$ as a function of $\beta^{XY}$. The dashed curves indicate the statistical error. The solid straight horizontal line gives $A_{3,2}^{BCSOS}$ at $K_R^{BCSOS}$. The vertical lines give the result for $\beta_{3,2}^{XY}$ and its error.

For the determination of $K_1^{SOS}$, we did not use $A_{1,l}$ directly, but the 'improved' quantity $D_{1,l}$: matching of the $A_{i,l}$-flows of two models happens if and only if also the $D_{i,l}$-flows match. This is so because for $B \to \infty$ the factors $A_{i,l}^{(0)}(B)$ converge to fixed points $A_i^{(0)}(B = \infty)$. This proves that we are allowed to use the $D_{i,l}$ instead of the $A_{i,l}$ without losing any control on the RG flow. Furthermore, the $D_{i,l}$ flows converge more rapidly than those of the $A_{i,l}$. This will be demonstrated below. Roughly speaking, the $A_{i,l}^{(0)}$ factors cancel irrelevant terms in the flow that anyway die out under successive RG steps. These terms are strong as long as $B$ is small, and are partly due to discretization details, e.g. of the lattice Laplacian. We want to stress the point that the use of the $D$'s instead of the $A$'s is by no means necessary: the changes on the larger lattices are negligible. However, using the $D's$ allows one to observe a collapse to a universal trajectory on much smaller lattices.

The solution of eq. (36) yields, for each $(L, l)$ pair, two values: $K_1^{SOS}$ and $K_3^{SOS}$. For an illustration of this first step, see fig. 5. Note that $K_1^{SOS}$ and $K_3^{SOS}$ will in general not be identical: one can expect matching only for a specific ratio $b_m^{SOS}$.

In a second step we plotted the values of $K_1^{SOS}$ and $K_3^{SOS}$ as a function of $B^{BCSOS}$. The couplings were linearly interpolated in $\log B^{BCSOS}$. The intersection of the two curves $K_1^{SOS}(B^{BCSOS})$ and $K_3^{SOS}(B^{BCSOS})$ then uniquely determines an estimate for the roughening coupling $K_R^{SOS}$ of the SOS model, see fig. 6.

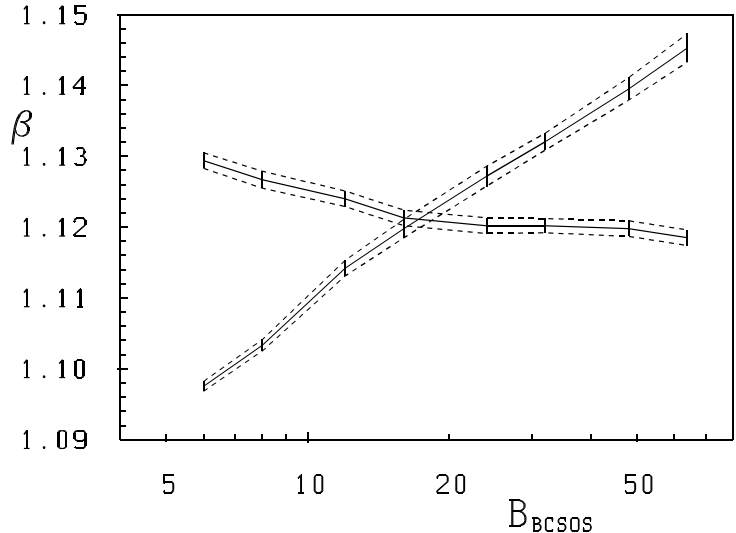

Figure 6: Determination of $\beta_R^{XY}$ for $L^{XY} = 32$ and $l = 2$. The curves give $\beta_1^{XY}$ and $\beta_3^{XY}$ as functions of $B_{BCSOS}$. The intersection of the curves uniquely determines $\beta_R^{XY}$ and the matching $B_{BCSOS}$.

In addition we obtain for each $B^{SOS}$ the BCSOS block size $B^{BCSOS}$ that leads to a matching in the sense described above. The results for the matching for the three models are summarized in table 6.

For the three models, the results for the roughening coupling $K_R$ obtained for the various lattice sizes $L$ and sizes $l$ of the blocked system are consistent with each other within statistical errors. Only the couplings for $l = 4$ on the smallest two lattice sizes and for $l = 2$ on the smallest lattice size deviate slightly from the rest. For the ratio of the matching block sizes $b_m = B^{SOS}/B^{BCSOS}$ the observation is the same. This indicates an extremely fast convergence to a universal RG flow of the models, since even for such small block sizes as $B^{BCSOS} = 16$ no deviation from the universal flow can be observed within our quite good statistics. To give estimates for the roughening coupling $K_R$ for the three models we averaged the values obtained for the largest $L$ with $l = 2$ and $l = 4$, and the second largest $L$ only with $l = 2$. We arrive at the following results

$$\begin{aligned} \beta_R^{XY} &= 1.1197(5)\,, \\ K_R^{DG} &= 0.6645(6)\,, \\ K_R^{ASOS} &= 0.8061(3)\,. \end{aligned} \qquad (37)$$

The quoted errors are statistical, but according to the discussion above, the systematic ones due to deviations from the universal parameter flow, should be much smaller. For the $b_m$ we find in a similar way

$$\begin{align}
b_m^{XY} &= 0.89(5)\,, \\
b_m^{DG} &= 0.31(2)\,, \\
b_m^{ASOS} &= 2.8(3)\,.
\end{align} \tag{38}$$

| Dual of XY model | | | | |
|---|---|---|---|---|
| $L$ | $\beta_R, l=2$ | $\beta_R, l=4$ | $b_m, l=2$ | $b_m, l=4$ |
| 16 | 1.1220(12) | 1.1257(8) | 0.84(5) | 0.75(1) |
| 24 | 1.1214(13) | 1.1225(8) | 0.91(7) | 0.84(2) |
| 32 | 1.1211(12) | 1.1214(8) | 0.93(10) | 0.85(3) |
| 48 | 1.1199(11) | 1.1205(7) | 0.89(13) | 0.89(3) |
| 64 | 1.1212(11) | 1.1201(7) | 0.89(5) | 0.82(12) |
| 96 | 1.1189(11) | 1.1194(7) | 0.89(12) | 0.95(7) |
| DG model | | | | |
| $L$ | $\beta_R, l=2$ | $\beta_R, l=4$ | $b_m, l=2$ | $b_m, l=4$ |
| 12 | 0.6627(16) | 0.6607(13) | 0.31(4) | 0.40(2) |
| 16 | 0.6650(13) | 0.6632(10) | 0.32(5) | 0.34(2) |
| 24 | 0.6633(16) | 0.6645(8) | 0.30(6) | 0.34(2) |
| 32 | 0.6650(16) | 0.6647(8) | 0.28(5) | 0.32(2) |
| ASOS model | | | | |
| $L$ | $\beta_R, l=2$ | $\beta_R, l=4$ | $b_m, l=2$ | $b_m, l=4$ |
| 32 | 0.8052(4) | | 2.3(2) | |
| 64 | 0.8061(6) | 0.8058(3) | 2.8(3) | 2.4(1) |
| 128 | 0.8061(6) | 0.8060(3) | 2.7(6) | 2.6(2) |
| 256 | 0.8060(5) | 0.8062(3) | 2.8(6) | 2.9(4) |

Table 6: $K_R$ and $b_m = B^{SOS}/B^{BCSOS}$ for the three SOS models as obtained from the matching of $A_{1,l}$ and $A_{3,l}$.

## 5.2 Demonstration that the matching is unique for all $A_{i,l}$

We want to demonstrate that all quantities $A_{i,l}$ ($D_{i,l}$) converge towards a universal flow with increasing $B^{SOS}$, provided that the couplings $K^{SOS}$ are tuned to their critical values quoted in eq. (37), and that the block size ratios $b_m^{SOS} = B^{SOS}/B^{BCSOS}$ are taken to be the matching values quoted in eq. (38). Recall that the critical couplings and $b_m$'s were determined by imposing the matching condition for $D_{1,l}$ and $A_{3,l}$ alone.

The first task was to evaluate the observables $A_{2,l}$, $A_{4,l}$ and $A_{5,l}$ at the critical couplings $K_R$ determined above using $D_{1,l}$ and $A_{3,l}$ only. The results are summarized in tables 7 to 9. For the sake of completeness we also give the values of the $D_{i,l}$, in table 10. We furthermore demonstrate the universal matching by plotting all measured block observables at criticality as functions of the matching block size $B^{BCSOS} = B^{SOS}/b_m^{SOS}$. The reader is invited to look carefully at figs. 7 to 23. To correctly interpret the plots, it is necessary to realize that the scale of the $y$-axis might differ from plot to plot. See, for example, figs. 7 and 8: The collapse of $D_{1,2}$ onto a universal curve is much faster than for the corresponding quantity $A_{1,2}$.

| $L$ | $l$ | $A_{1,l}$ | $A_{2,l}$ | $A_{3,l}$ | $A_{4,l}$ | $A_{5,l}$ |
|---|---|---|---|---|---|---|
| 16 | 1 | | | 0.2601(19) | 0.0669(17) | 0.0175(13) |
| 24 | 1 | | | 0.2381(28) | 0.0497(20) | 0.0110(17) |
| 32 | 1 | | | 0.2260(22) | 0.0493(18) | 0.0113(16) |
| 48 | 1 | | | 0.2029(24) | 0.0391(21) | 0.0084(18) |
| 64 | 1 | | | 0.1946(31) | 0.0354(19) | 0.0072(19) |
| 96 | 1 | | | 0.1748(31) | 0.0256(26) | 0.0032(18) |
| 16 | 2 | 0.08486(17) | 0.11735(30) | 0.2187(13) | 0.0531(8) | 0.0140(7) |
| 24 | 2 | 0.08255(20) | 0.11505(32) | 0.2000(16) | 0.0436(10) | 0.0101(7) |
| 32 | 2 | 0.08146(20) | 0.11373(32) | 0.1893(13) | 0.0377(10) | 0.0088(8) |
| 48 | 2 | 0.08109(23) | 0.11337(37) | 0.1712(14) | 0.0306(10) | 0.0071(9) |
| 64 | 2 | 0.08051(23) | 0.11234(35) | 0.1629(18) | 0.0266(8) | 0.0026(9) |
| 96 | 2 | 0.08087(26) | 0.11337(44) | 0.1470(20) | 0.0207(11) | 0.0038(10) |
| 16 | 4 | 0.17102(20) | 0.22671(30) | 0.2088(6) | 0.0624(5) | 0.0240(4) |
| 24 | 4 | 0.16283(18) | 0.21812(28) | 0.1883(6) | 0.0479(5) | 0.0144(3) |
| 32 | 4 | 0.15965(18) | 0.21477(26) | 0.1755(7) | 0.0414(5) | 0.0107(4) |
| 48 | 4 | 0.15721(16) | 0.21206(23) | 0.1602(7) | 0.0338(5) | 0.0076(5) |
| 64 | 4 | 0.15602(20) | 0.21065(29) | 0.1489(8) | 0.0283(6) | 0.0071(5) |
| 96 | 4 | 0.15531(20) | 0.21012(30) | 0.1377(9) | 0.0234(6) | 0.0050(6) |

Table 7: Finite lattice renormalization group flow of the $A_{i,l}$ for the dual of the XY model at $\beta = 1.1197$

| $L$ | $l$ | $A_{1,l}$ | $A_{2,l}$ | $A_{3,l}$ | $A_{4,l}$ | $A_{5,l}$ |
|---|---|---|---|---|---|---|
| 12 | 1 | | | 0.2170(31) | 0.0474(22) | 0.0125(17) |
| 16 | 1 | | | 0.2061(23) | 0.0403(27) | 0.0068(15) |
| 24 | 1 | | | 0.1944(45) | 0.0407(32) | 0.0057(25) |
| 32 | 1 | | | 0.1729(41) | 0.0312(29) | 0.0058(27) |
| 12 | 2 | 0.08611(30) | 0.11901(45) | 0.1850(18) | 0.0409(12) | 0.0090(10) |
| 16 | 2 | 0.08408(27) | 0.11711(40) | 0.1725(16) | 0.0314(11) | 0.0062(12) |
| 24 | 2 | 0.08185(38) | 0.11418(58) | 0.1603(27) | 0.0275(14) | 0.0064(11) |
| 32 | 2 | 0.08140(34) | 0.11431(59) | 0.1458(27) | 0.0246(15) | 0.0051(13) |
| 12 | 4 | 0.17782(32) | 0.23433(45) | 0.1875(8) | 0.0588(6) | 0.0327(6) |
| 16 | 4 | 0.16839(24) | 0.22388(35) | 0.1663(8) | 0.0413(6) | 0.0143(6) |
| 24 | 4 | 0.16100(26) | 0.21601(40) | 0.1488(10) | 0.0305(8) | 0.0068(6) |
| 32 | 4 | 0.15800(25) | 0.21312(37) | 0.1380(10) | 0.0254(7) | 0.0050(5) |

Table 8: Finite lattice renormalization group flow of the $A_{i,l}$ for the DG model at $K^{DG} = 0.6645$

| $L$ | $l$ | $A_{1,l}$ | $A_{2,l}$ | $A_{3,l}$ | $A_{4,l}$ | $A_{5,l}$ |
|---|---|---|---|---|---|---|
| 32 | 1 | | | 0.2800(31) | 0.0766(27) | 0.0232(19) |
| 64 | 1 | | | 0.2404(35) | 0.0525(27) | 0.0132(25) |
| 128 | 1 | | | 0.2103(41) | 0.0425(28) | 0.0084(24) |
| 256 | 1 | | | 0.1866(45) | 0.0312(28) | 0.0028(24) |
| 32 | 2 | 0.08291(30) | 0.11513(52) | 0.2368(18) | 0.0609(12) | 0.0191(11) |
| 64 | 2 | 0.08202(33) | 0.11504(46) | 0.2023(25) | 0.0424(13) | 0.0088(12) |
| 128 | 2 | 0.08083(34) | 0.11337(52) | 0.1756(25) | 0.0334(14) | 0.0067(11) |
| 256 | 2 | 0.08069(33) | 0.11313(57) | 0.1553(26) | 0.0251(15) | 0.0035(13) |
| 32 | 4 | 0.16406(29) | 0.21936(45) | 0.2230(9) | 0.0656(7) | 0.0222(7) |
| 64 | 4 | 0.15914(27) | 0.21457(37) | 0.1891(9) | 0.0471(7) | 0.0136(7) |
| 128 | 4 | 0.15651(28) | 0.21137(43) | 0.1635(13) | 0.0320(7) | 0.0072(6) |
| 256 | 4 | 0.15513(28) | 0.21010(42) | 0.1412(11) | 0.0255(9) | 0.0057(7) |

Table 9: Finite lattice renormalization group flow of the $A_{i,l}$ for the ASOS model at $K^{ASOS} = 0.8061$

| BCSOS model | | | | | |
|---|---|---|---|---|---|
| $B$ | $D_{1,2}$ | $D_{2,2}$ | $B$ | $D_{1,4}$ | $D_{2,4}$ |
| 6 | 0.08313(15) | 0.11352(25) | 3 | 0.17163(15) | 0.22080(22) |
| 8 | 0.08244(21) | 0.11389(35) | 4 | 0.16395(19) | 0.21711(29) |
| 12 | 0.08175(22) | 0.11347(36) | 6 | 0.15955(19) | 0.21389(29) |
| 16 | 0.08106(23) | 0.11307(39) | 8 | 0.15770(21) | 0.21210(31) |
| 24 | 0.08078(22) | 0.11304(37) | 12 | 0.15658(18) | 0.21132(28) |
| 32 | 0.08077(17) | 0.11317(29) | 16 | 0.15567(14) | 0.21054(21) |
| 48 | 0.08067(20) | 0.11316(33) | 24 | 0.15487(16) | 0.20959(25) |
| 64 | 0.08033(21) | 0.11263(35) | 32 | 0.15451(17) | 0.20908(27) |
| Dual of XY model | | | | | |
| $B$ | $D_{1,2}$ | $D_{2,2}$ | $B$ | $D_{1,4}$ | $D_{2,4}$ |
| 8 | 0.08164(16) | 0.11379(29) | 4 | 0.15854(19) | 0.21358(28) |
| 12 | 0.08112(20) | 0.11347(32) | 6 | 0.15725(17) | 0.21228(27) |
| 16 | 0.08066(20) | 0.11285(32) | 8 | 0.15651(18) | 0.21149(26) |
| 24 | 0.08073(23) | 0.11298(37) | 12 | 0.15581(16) | 0.21060(23) |
| 32 | 0.08031(23) | 0.11212(35) | 16 | 0.15523(20) | 0.20983(29) |
| 48 | 0.08078(26) | 0.11327(44) | 24 | 0.15496(20) | 0.20976(30) |
| DG model | | | | | |
| $B$ | $D_{1,2}$ | $D_{2,2}$ | $B$ | $D_{1,4}$ | $D_{2,4}$ |
| 6 | 0.08051(28) | 0.11275(43) | 3 | 0.15632(28) | 0.21143(41) |
| 8 | 0.08089(26) | 0.11356(39) | 4 | 0.15610(22) | 0.21092(33) |
| 12 | 0.08043(37) | 0.11262(57) | 6 | 0.15548(25) | 0.21022(39) |
| 16 | 0.08060(34) | 0.11342(59) | 8 | 0.15489(25) | 0.20986(36) |
| ASOS model | | | | | |
| $B$ | $D_{1,2}$ | $D_{2,2}$ | $B$ | $D_{1,4}$ | $D_{2,4}$ |
| 16 | 0.08209(30) | 0.11424(52) | 8 | 0.16083(28) | 0.21601(44) |
| 32 | 0.08182(33) | 0.11482(46) | 16 | 0.15834(27) | 0.21374(37) |
| 64 | 0.08078(34) | 0.11331(52) | 32 | 0.15631(28) | 0.21116(43) |
| 128 | 0.08068(33) | 0.11312(57) | 64 | 0.15508(28) | 0.21005(42) |

Table 10: Results for the flow of the $D_{i,l}$ at criticality

## 5.3 Determination of non-universal constants

The matching also allows us to determine the non-universal constants appearing in the formulae for the divergence of observables near the roughening transition. Let us discuss this over the example of the correlation length $\xi$. Its critical behaviour is

$$\xi \simeq A \, \exp\left(C \kappa^{-1/2}\right), \quad \kappa = \frac{K - K_R}{K_R}. \tag{39}$$

Let us consider matching on an RG trajectory in the phase with finite correlation length (smooth phase), close to the critical trajectory. Let us assume that the BCSOS block observables match the SOS block observables for sufficiently large $B^{BCSOS}$, $B^{SOS}$, with $B^{SOS} = b_m^{SOS} B^{BCSOS}$. Then

$$\xi^{SOS} = b_m^{SOS} \xi^{BCSOS}. \tag{40}$$

Inserting eq. (39) in eq. (40) we get

$$\xi^{SOS} \simeq b_m^{SOS} A^{BCSOS} \exp\left(C^{BCSOS} \left(\kappa^{BCSOS}\right)^{-1/2}\right). \tag{41}$$

It is a general assumption of the renormalization group that couplings on the blocked system are smooth functions of the block size and the coupling on the fine lattice. We thus assume that $b_m^{SOS}$ is a smooth function of $\kappa^{SOS}$, even at the roughening transition,

$$b_m^{SOS}(K) = b_m^{SOS}(K_R) + O\left(\kappa^{SOS}\right). \tag{42}$$

Furthermore, also $\kappa^{BCSOS}$ is a smooth function of $\kappa^{SOS}$,

$$\kappa^{BCSOS} = q \, \kappa^{SOS} + O\left(\left(\kappa^{SOS}\right)^2\right). \tag{43}$$

In the limit $\kappa^{SOS} \to 0$ we get

$$\xi^{SOS} \simeq A^{SOS} \exp\left(C^{SOS}(\kappa^{SOS})^{-1/2}\right), \tag{44}$$

with

$$\begin{aligned} A^{SOS} &= b_m^{SOS} A^{BCSOS}, \\ C^{SOS} &= q^{-1/2} C^{BCSOS}. \end{aligned} \tag{45}$$

With the results obtained above for $b_m$ and with $A^{BCSOS} = \frac{1}{4}$ we find

$$\begin{aligned} A^{XY} &= 0.223(13), \\ A^{DG} &= 0.078(5), \\ A^{ASOS} &= 0.70(8). \end{aligned} \tag{46}$$

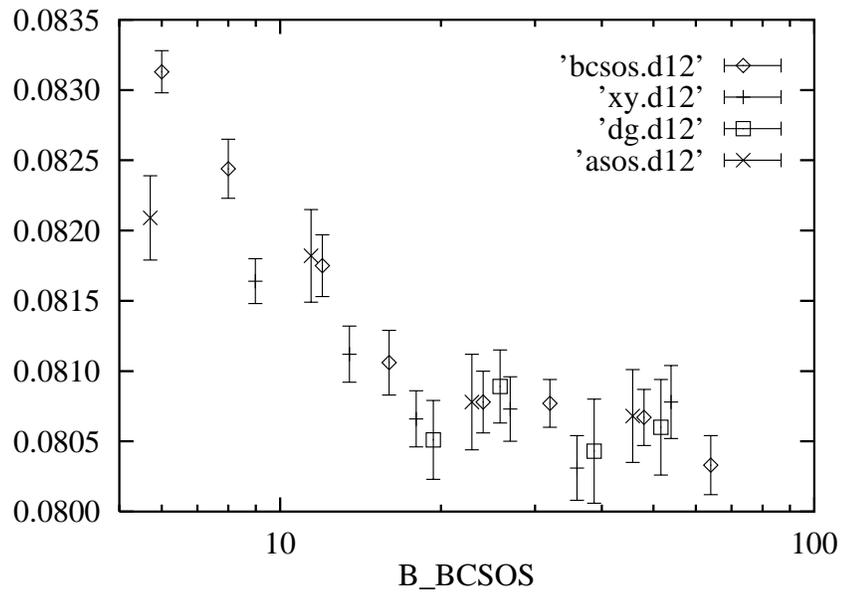

Figure 7: $D_{1,2}$ at criticality, plotted as a function of $B^{BCSOS} = B^{SOS}/b_m^{SOS}$

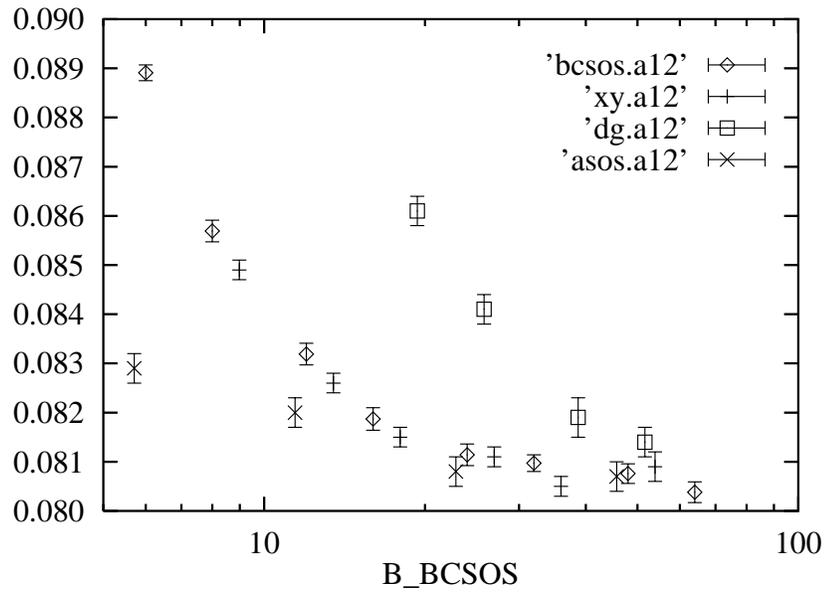

Figure 8: $A_{1,2}$ at criticality, plotted as a function of $B^{BCSOS} = B^{SOS}/b_m^{SOS}$

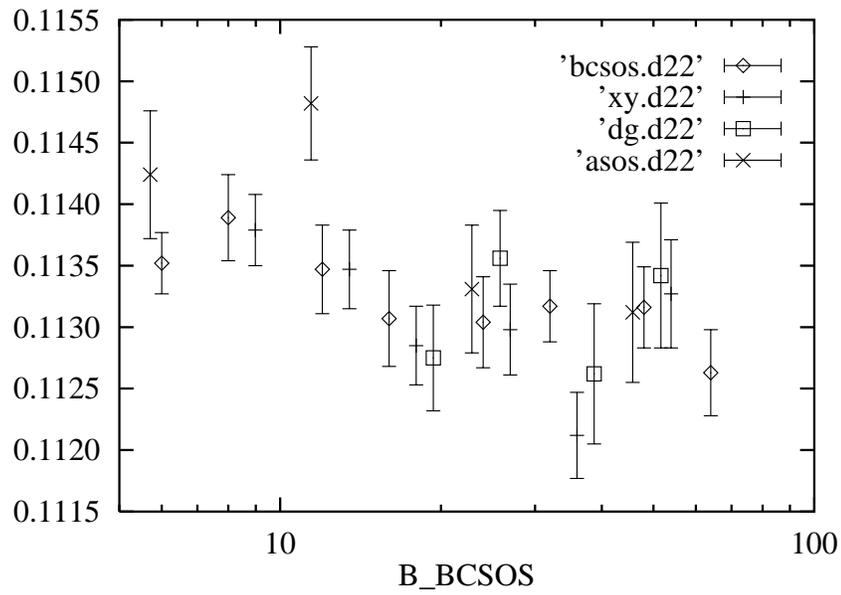

Figure 9: $D_{2,2}$ at criticality, plotted as a function of $B^{BCSOS} = B^{SOS}/b_m^{SOS}$

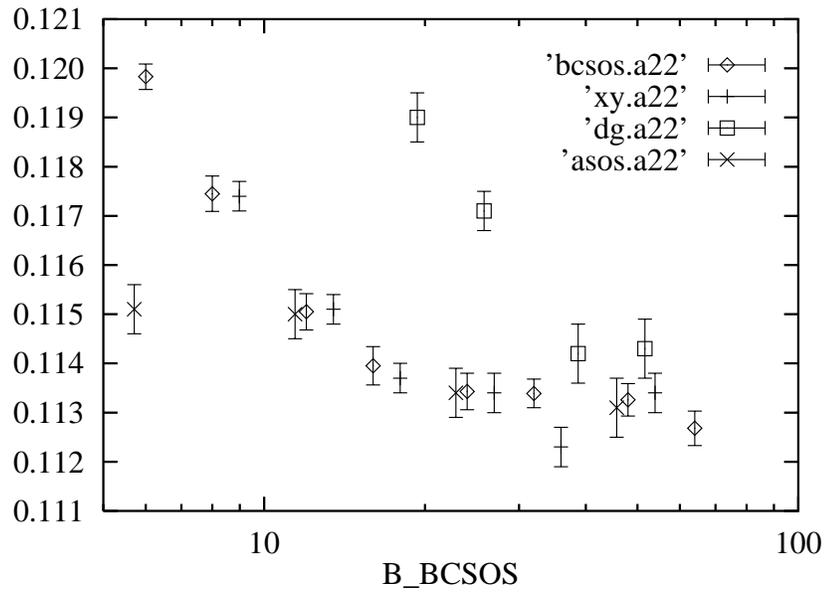

Figure 10: $A_{2,2}$ at criticality, plotted as a function of $B^{BCSOS} = B^{SOS}/b_m^{SOS}$

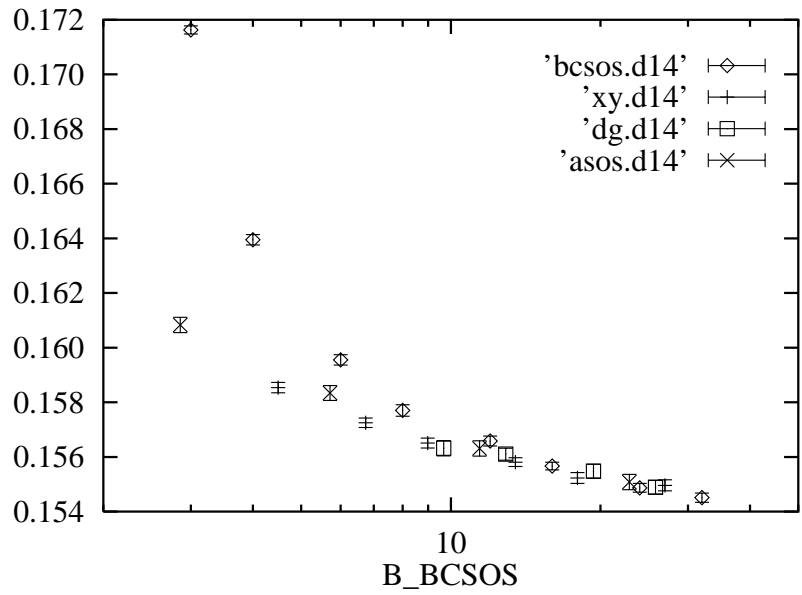

Figure 11: $D_{1,4}$ at criticality, plotted as a function of $B^{BCSOS} = B^{SOS}/b_m^{SOS}$

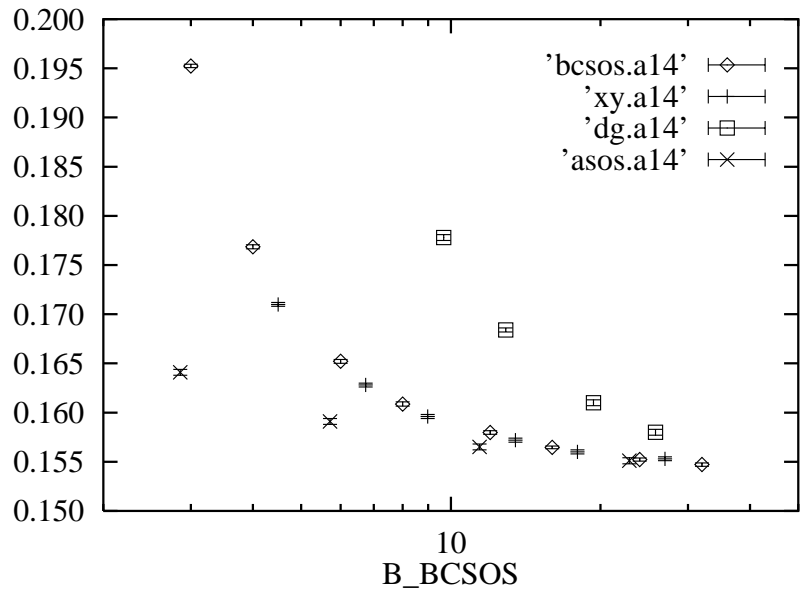

Figure 12: $A_{1,4}$ at criticality, plotted as a function of $B^{BCSOS} = B^{SOS}/b_m^{SOS}$

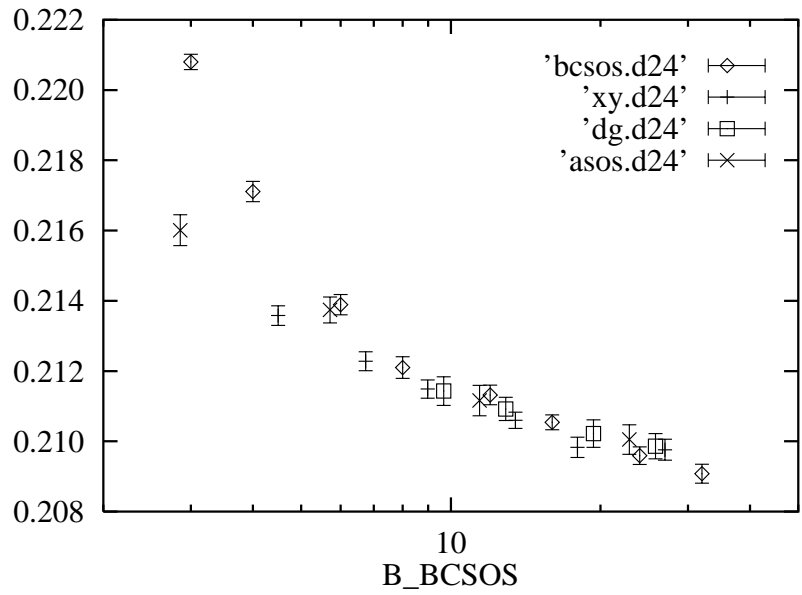

Figure 13: $D_{2,4}$ at criticality, plotted as a function of $B^{BCSOS} = B^{SOS}/b_m^{SOS}$

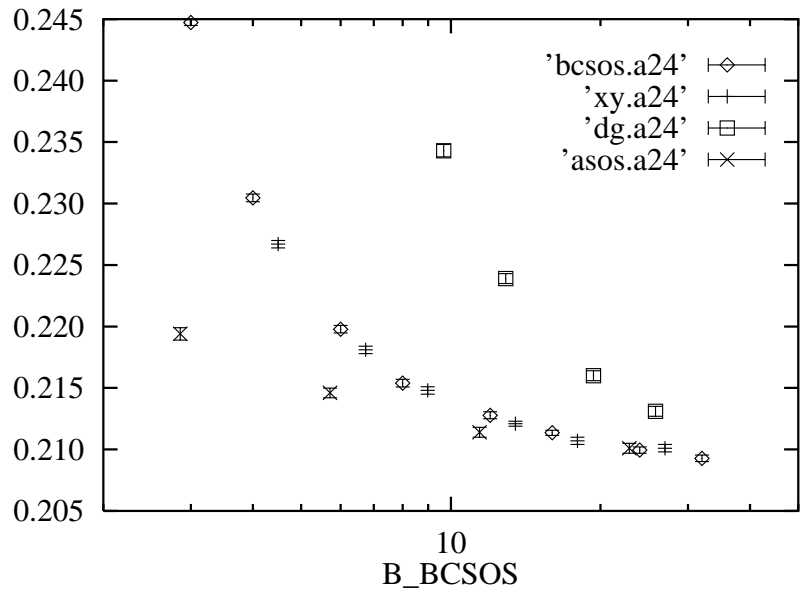

Figure 14: $A_{2,4}$ at criticality, plotted as a function of $B^{BCSOS} = B^{SOS}/b_m^{SOS}$

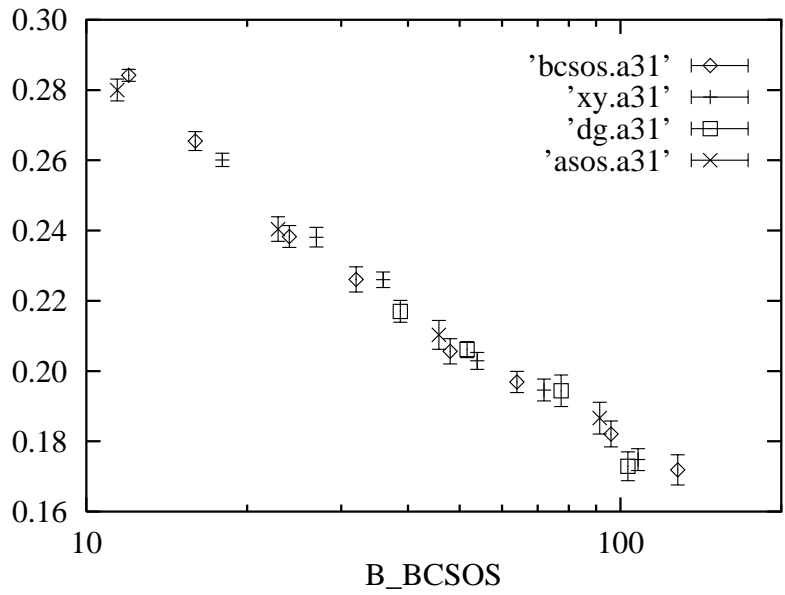

Figure 15: $A_{3,1}$ at criticality, plotted as a function of $B^{BCSOS} = B^{SOS}/b_m^{SOS}$

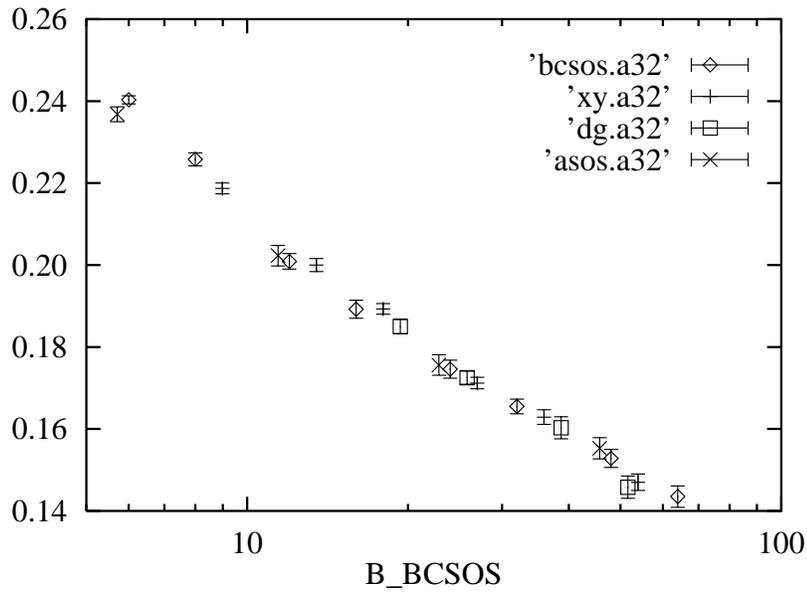

Figure 16: $A_{3,2}$ at criticality, plotted as a function of $B^{BCSOS} = B^{SOS}/b_m^{SOS}$

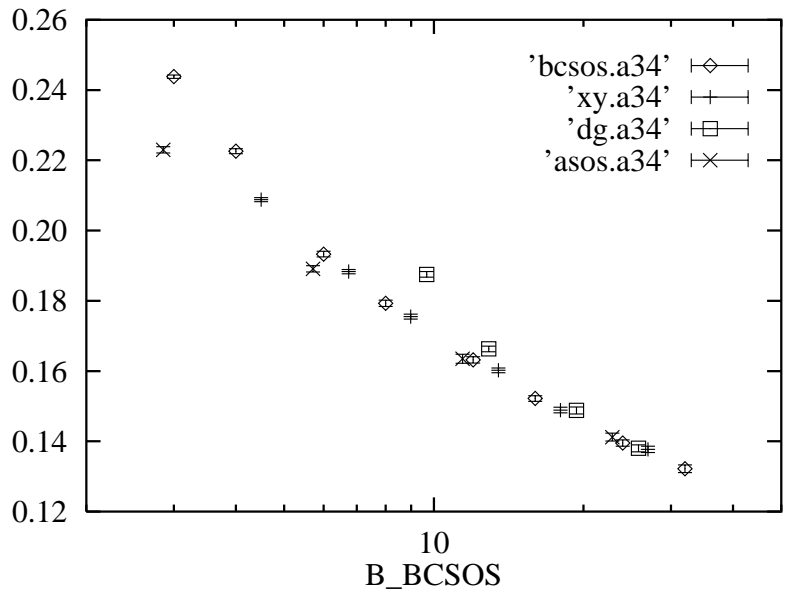

Figure 17: $A_{3,4}$ at criticality, plotted as a function of $B^{BCSOS} = B^{SOS}/b_m^{SOS}$

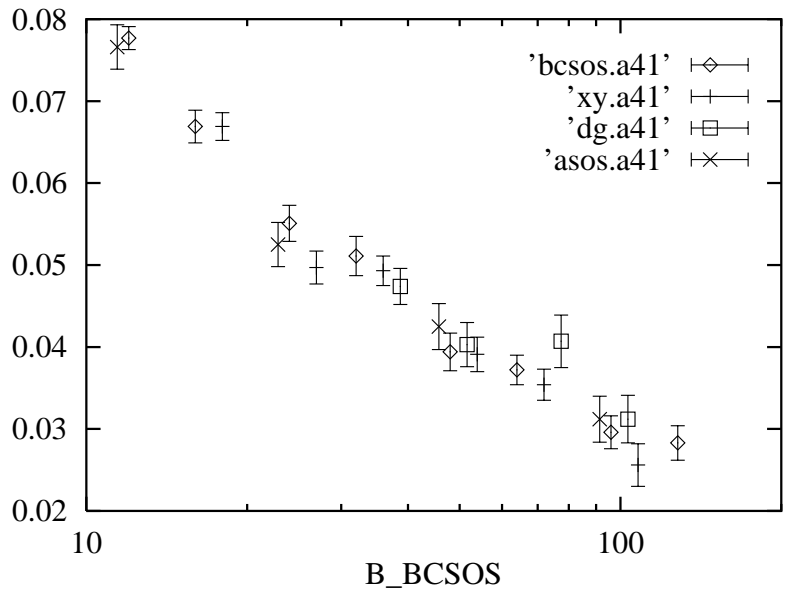

Figure 18: $A_{4,1}$ at criticality, plotted as a function of $B^{BCSOS} = B^{SOS}/b_m^{SOS}$

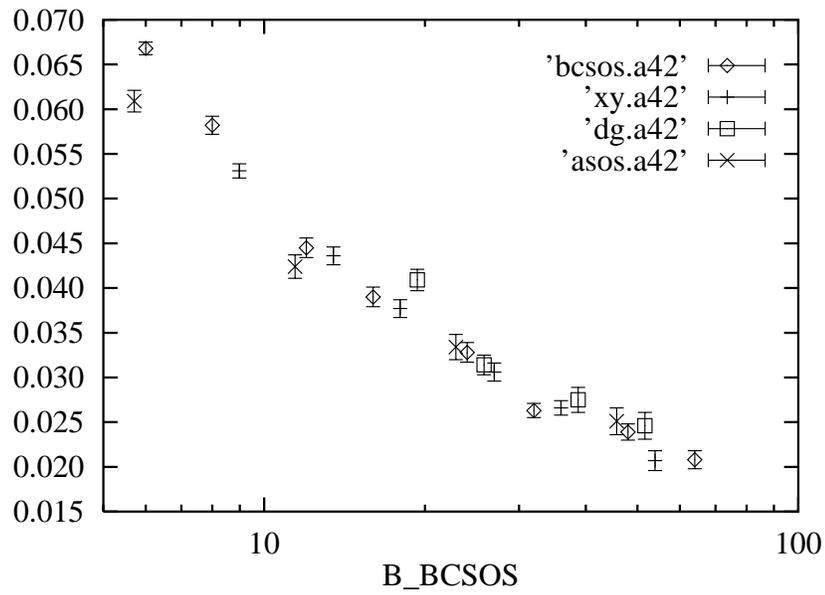

Figure 19: $A_{4,2}$ at criticality, plotted as a function of $B^{BCSOS} = B^{SOS}/b_m^{SOS}$

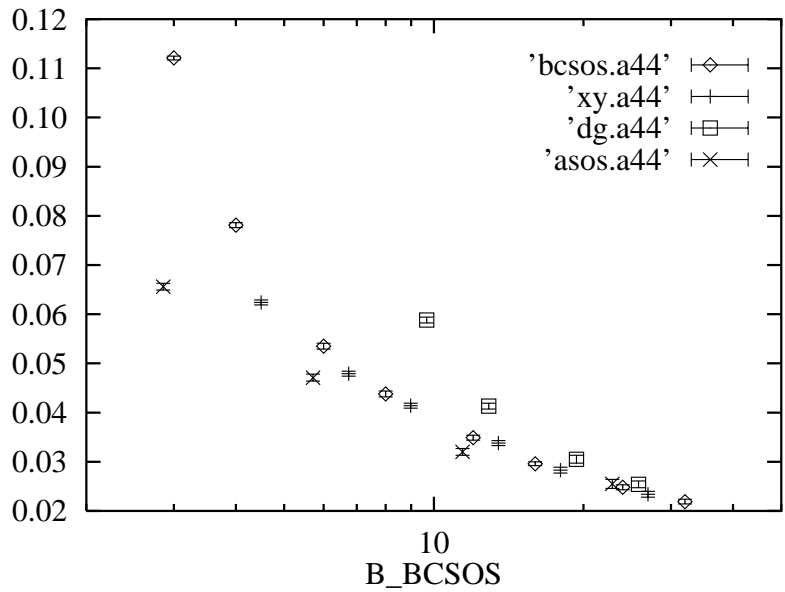

Figure 20: $A_{4,4}$ at criticality, plotted as a function of $B^{BCSOS} = B^{SOS}/b_m^{SOS}$

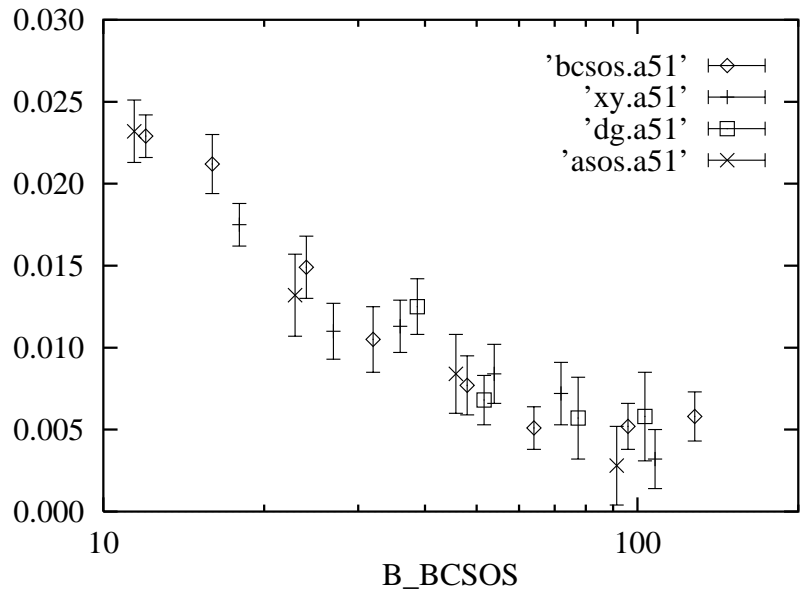

Figure 21: $A_{5,1}$ at criticality, plotted as a function of $B^{BCSOS} = B^{SOS}/b_m^{SOS}$

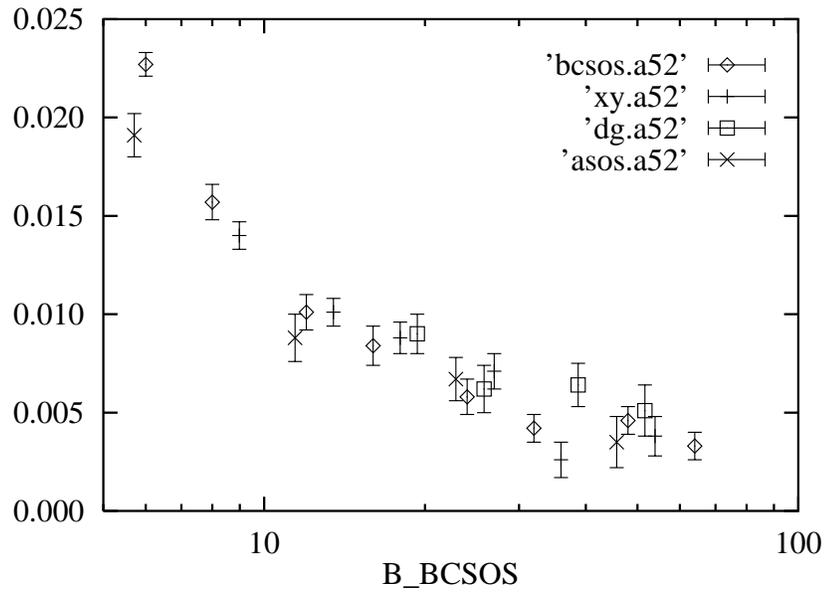

Figure 22: $A_{5,2}$ at criticality, plotted as a function of $B^{BCSOS} = B^{SOS}/b_m^{SOS}$

There remains to determine the constant $q$ connecting $\kappa^{SOS}$ and $\kappa^{BCSOS}$. For this purpose let us rewrite the matching condition for coupling constants in the neighbourhood of the critical point. In order to keep the formulae compact, we shall write an '$S$' for 'SOS' and a '$B$' for 'BCSOS'. The matching condition reads

$$A_{i,l}^S\left(B^S, K_R^S + k^S\right) = A_{i,l}^B\left(B^B, K_R^B + k^B\right) . \tag{47}$$

A Taylor expansion around $K_R^S$ and $K_R^B$, respectively, yields

$$A_{i,l}^S\left(B^S, K_R^S\right) + k^S \left(\frac{\partial A_{i,l}^S}{\partial K^S}\left(B^S, K_R^S\right) + \frac{\partial A_{i,l}^S}{\partial B^S}\left(B^S, K_R^S\right)\frac{\partial B^S}{\partial K^S}\left(K_R^S\right)\right)$$
$$= \text{(same for BCSOS)} . \tag{48}$$

These equations simplify since the observables match at criticality. As a consequence of the matching condition,

$$A_{i,l}^S\left(B^S, K_R^S\right) \equiv A_{i,l}^B\left(B^B, K_R^B\right) ,$$
$$\frac{\partial B^S}{\partial K^S}\left(K_R^S\right) \equiv 0 . \tag{49}$$

The second of these equations expresses the fact that we keep $B^S$ fixed when tuning the other parameters in order to fulfil the matching condition. We are left with

$$k^S \left(\frac{\partial A_{i,l}^S}{\partial K^S}\left(B^S, K_R^S\right)\right) = k^B \left(\frac{\partial A_{i,l}^B}{\partial K^B}\left(B^B, K_R^B\right) + \underbrace{\frac{\partial A_{i,l}^B}{\partial B^B}\left(B^B, K_R^B\right)\frac{\partial B^B}{\partial K^B}(K_R^B)}_{d_{i,l}}\right) . \tag{50}$$

Let us now again restrict our attention to $i = 1, 3$. We can then solve eqs. (50) with respect to $k^S/k^B$:

$$\frac{k^S}{k^B} = \frac{\frac{\partial A_{1,l}^B}{\partial K^B}/\frac{\partial A_{1,l}^B}{\partial B^B} - \frac{\partial A_{3,l}^B}{\partial K^B}/\frac{\partial A_{3,l}^B}{\partial B^B}}{\frac{\partial A_{1,l}^S}{\partial K^S}/\frac{\partial A_{1,l}^B}{\partial B^B} - \frac{\partial A_{3,l}^S}{\partial K^S}/\frac{\partial A_{3,l}^B}{\partial B^B}} . \tag{51}$$

The partial derivatives $\partial A/\partial K$ can be determined with reweighting methods or with the help of the formula

$$\frac{\partial A}{\partial K} = -\langle AH\rangle + \langle A\rangle\langle H\rangle . \tag{52}$$

$H$ is the Hamiltonian, and the expectation values are taken in the system with partition function $Z = \sum_{\text{conf}} \exp(-KH)$. The partial derivatives $\partial A_{i,l}^B/\partial B^B$ can be extracted from table 5.

However, eq. (51) simplifies very much if the term $d_{i,l}$ in eq. (50) can be neglected. This is the case when

$$\frac{\partial B^B}{\partial K^B}(K_R^B) \approx 0 . \tag{53}$$

| Dual of XY model | | | | | |
|---|---|---|---|---|---|
| $L$ | $i=3, l=1$ | $i=1, l=2$ | $i=3, l=2$ | $i=1, l=4$ | $i=3, l=4$ |
| 16 | 0.413(8) | 0.430(11) | 0.406(8) | 0.424(10) | 0.400(7) |
| 24 | 0.413(9) | 0.436(14) | 0.419(8) | 0.431(14) | 0.418(8) |
| 32 | 0.427(12) | 0.461(19) | 0.429(10) | 0.437(19) | 0.421(9) |
| 48 | 0.434(15) | 0.441(21) | 0.423(12) | 0.445(22) | 0.445(9) |
| 64 | 0.449(17) | 0.420(28) | 0.445(13) | 0.447(29) | 0.434(11) |
| 96 | 0.449(27) | 0.447(46) | 0.440(21) | 0.437(45) | 0.424(15) |
| Discrete Gaussian model | | | | | |
| $L$ | $i=3, l=1$ | $i=1, l=2$ | $i=3, l=2$ | $i=1, l=4$ | $i=3, l=4$ |
| 12 | 0.381(12) | 0.394(17) | 0.375(10) | 0.394(18) | 0.369(9) |
| 16 | 0.381(13) | 0.384(18) | 0.379(11) | 0.385(19) | 0.374(8) |
| 24 | 0.388(17) | 0.384(30) | 0.381(13) | 0.388(29) | 0.374(11) |
| 32 | 0.397(24) | 0.364(38) | 0.394(19) | 0.372(39) | 0.387(13) |
| ASOS model | | | | | |
| $L$ | $i=3, l=1$ | $i=1, l=2$ | $i=3, l=2$ | $i=1, l=4$ | $i=3, l=4$ |
| 32 | 1.482(32) | 1.504(35) | 1.451(25) | 1.426(40) | 1.345(21) |
| 64 | 1.385(42) | 1.474(46) | 1.345(40) | 1.457(44) | 1.447(42) |
| 128 | 1.383(50) | 1.416(66) | 1.373(43) | 1.510(72) | 1.421(41) |
| 256 | 1.909(106) | 1.625(157) | 1.712(75) | 1.608(151) | 1.586(57) |

Table 11: $(k^S/k^B)^{\text{naive}}_{i,l}$ for the three SOS models

Equivalently, the simplification relies on the assumption that the expansion of the matching $B^B$ around the roughening coupling $K_R^B$ is of second order in $k^B$. Then we get

$$\left(\frac{k^S}{k^B}\right)^{\text{naive}}_{i,l} = \frac{\partial A^B_{i,l}}{\partial K^B} \bigg/ \frac{\partial A^S_{i,l}}{\partial K^S} , \quad i = 1, 3 . \tag{54}$$

If the approximation is a good one, the $(k^S/k^B)^{\text{naive}}_{i,l}$ should be independent of $i, l$. Our results for these quantities are summarized in table 11.

For all three models the numbers for $(k^S/k^B)^{\text{naive}}_{i,l}$ for different $(i, l)$ are consistent within the error-bars. We arrive at the following estimates:

$$\begin{aligned} k^{BCSOS}/k^{XY} &= 0.43(1) , \\ k^{BCSOS}/k^{DG} &= 0.39(1) , \\ k^{BCSOS}/k^{ASOS} &= 1.46(6) . \end{aligned} \tag{55}$$

Now we use that

$$C^{SOS} = \left(\frac{\kappa^{SOS}}{\kappa^{BCSOS}}\right)^{1/2} C^{BCSOS} \qquad (56)$$

and $\kappa^{SOS} = k^{SOS}/K_R^{SOS}$. We get

$$\begin{aligned} C^{XY} &= 1.78(2), \\ C^{DG} &= 2.44(3), \\ C^{ASOS} &= 1.14(2). \end{aligned} \qquad (57)$$

### 5.4  Comparison with other numerical studies

We compared our results with those obtained in other Monte Carlo studies.

Janke and Nather [51] simulated the XY model with the Villain action in the vortex phase using Wolff's single cluster algorithm [60]. They measured correlation lengths up to $\xi = 140$ on lattices up to $L = 1200$, at $\beta$ ranging from $\beta^V = 0.590$ up to $\beta^V = 0.675$, where $\beta^V = 0.5/K^{DG}$. They fitted their results for $\xi$ to eq. (39). To check for systematical errors due to a too large distance to the critical point, they used two different definitions of $\kappa$:

$$\begin{aligned} \kappa_T &= |T - T_c|/T_c, \\ \kappa_\beta &= |\beta - \beta_c|/\beta_c. \end{aligned} \qquad (58)$$

The two definitions agree to the first-order Taylor expansion around the critical point. Hence both fits should give consistent results when the data included are obtained in a sufficiently small neighbourhood of the critical point. The comparison of the results of Janke and Nather with our results is given in table 12. The results of the two fits are not consistent within the error-bars. One is therefore led to the conclusion that the systematic error due to a too large distance of the simulation points from criticality is much larger than the quoted statistical errors.

The authors give $\beta_c = 0.752(5)$ as an overall estimate for the critical coupling. Taking into account the systematic errors of the fits, the results of their simulation are well consistent with our results.

In [30], the roughening coupling of the DG model was estimated from fits of the finite size behaviour of the surface thickness. The fits were done with a renormalization group improved formula. The best estimate was $\beta_c = 0.5/K_R = 0.755(3)$, which is nicely consistent with the estimate arrived at in the present paper, namely $\beta_c = 0.7524(7)$.

In the case of the XY model with cosine action we can compare our results with a fit given in [75], which includes data of [50] and data of the authors, see table 13. As in the case of the DG model one can say that the results compare well with ours taking the systematic errors of the fits into account. We also include the results of the MCRG study [52] in this comparison.

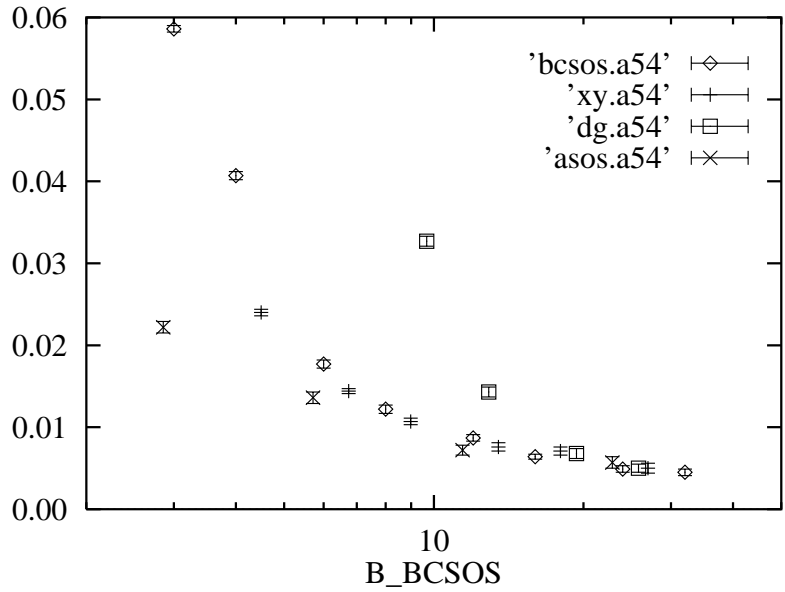

Figure 23: $A_{5,4}$ at criticality, plotted as a function of $B^{BCSOS} = B^{SOS}/b_m^{SOS}$

| Definition of $\kappa$ | $\beta_c^V$ | $A$ | $C$ |
|---|---|---|---|
| $\kappa_T$ | 0.75106(36) | 0.1204(18) | 2.370(11) |
| $\kappa_\beta$ | 0.75814(40) | 0.0287(7) | 2.812(14) |
| This work | 0.7524(7) | 0.078(5) | 2.44(3) |

Table 12: Comparison of our results for the DG model with those of ref. [51]. The relation between $\beta^V$ and $K^{DG}$ is $\beta^V = 0.5/K^{DG}$

| Authors | $\beta_c$ | $A$ | $C$ |
|---|---|---|---|
| Gupta et al. | 1.1218 | 0.2129 | 1.7258 |
| Biferale | 1.112(2) | | 1.74(20) |
| This work | 1.1197(5) | 0.223(13) | 1.78(2) |

Table 13: Comparison of our results for the XY model with those of refs. [75] and [52]

# Appendix 1: Solution of the KT equations

In this appendix, we will solve the KT equations

$$\begin{aligned} \dot{y} &= -xy, \\ \dot{x} &= -y^2. \end{aligned} \tag{59}$$

It is easy to see that $dy^2/dt = dx^2/dt$. Therefore $E = y^2 - x^2$ is a constant of motion. We consider separately the three cases $E = 0$, $E < 0$, and $E > 0$.

$E = 0$: From $y^2 - x^2 = 0$ it follows that either $x = y$ or $x = -y$. In the first case one is on the critical trajectory that runs into the fixed point at $(x, y) = (0, 0)$. The second case corresponds to the expanding manifold that leaves the fixed point. On the critical trajectory $y$ obeys the differential equation $\dot{y} = -y^2$, which has the solution

$$y(t) = \frac{1}{\frac{1}{y_0} + t}, \tag{60}$$

where $y_0 = y(t = 0)$. On the expanding manifold $y$ obeys $\dot{y} = y^2$, and the solution is

$$y(t) = \frac{1}{\frac{1}{y_0} - t}. \tag{61}$$

Note that on this trajectory $y$ diverges for finite $t$.

$E < 0$: We look at the solutions in the $x > 0$ region (the solutions for $x < 0$ can be obtained analogously). Let $a$ denote the point where the trajectory hits the $x$-axis at infinite $t$. Define $x = a + f$. We then have the system of differential equations

$$\begin{aligned} \frac{dy^2}{dt} &= -2(a+f)y^2, \\ \frac{df}{dt} &= -y^2. \end{aligned} \tag{62}$$

It follows that

$$\frac{dy^2}{df} = 2(a+f), \tag{63}$$

and therefore

$$y^2 = f^2 + 2af. \tag{64}$$

The integration constant vanishes because $y = 0$ for $f = 0$. We find the following differential equation for $f$:

$$\frac{df}{dt} = -f(f + 2a). \tag{65}$$

This equation can be solved by separation of variables

$$dt = -\frac{df}{2a}\left(\frac{1}{f} - \frac{1}{f + 2a}\right). \tag{66}$$

By integration we find

$$f(t) = 2a \left\{ \left(1 + \frac{2a}{f(t_0)}\right) e^{2a(t-t_0)} - 1 \right\}^{-1} . \tag{67}$$

$E > 0$: Define $\epsilon = \sqrt{E}$. From $y^2 = \epsilon^2 + x^2$ one obtains

$$\frac{dx}{dt} = -(\epsilon^2 + x^2) . \tag{68}$$

This differential equation can again be solved by separation of variables. The solution is

$$x(t) = \epsilon \tan\left(\arctan\frac{x(t_0)}{\epsilon} - \epsilon(t - t_0)\right) . \tag{69}$$

## Appendix 2: Gaussian model block correlation functions

In this appendix, we shall derive an explicit formula for the block correlation functions of the massless Gaussian model defined by the Hamiltonian

$$\mathcal{H}(\varphi) = \tfrac{1}{2} (\varphi, -\Delta\varphi) = \sum_{<x,y>} (\varphi_x - \varphi_y)^2 . \tag{70}$$

The block lattice $\Lambda'$ consists of $l \times l$ blocks $x'$. Each block $x'$ is built of $B \times B$ fine grid points $x$, so that the fine lattice has extension $L \times L$ with $L = lB$. All lattices involved are provided with periodic boundary conditions. We define a block-block correlation function by

$$\begin{aligned} G(x', y') &= <(\phi_{x'} - \phi_{y'})^2> , \\ \phi_{x'} &= B^{-2} \sum_{x \in x'} \varphi_x . \end{aligned} \tag{71}$$

$G$ has the following Fourier representation:

$$G(x', y') = 2B^{-4}L^{-2} \sum_{p \neq 0} \hat{p}^{-2} \left( \sum_{x \in x'} \sum_{y \in x'} \exp(ip(x-y)) - \sum_{x \in x'} \sum_{y \in y'} \exp(ip(x-y)) \right) . \tag{72}$$

The periodic momentum squared is defined through

$$\hat{p}^2 = 4 - 2\cos p_1 - 2\cos p_2 , \tag{73}$$

where the $p_i$ denote components of the lattice momentum $p$. The $p_i$ run over the values $(0, 1, ..., L-1)\frac{2\pi}{L}$. It is not difficult to derive that

$$\sum_{x \in x'} \sum_{y \in y'} \exp(ip(x-y)) = F(p_1) F(p_2) \exp(ipB(x'-y')) , \tag{74}$$

with

$$F(p_i) = \begin{cases} B^2 & \text{if } p_i = 0, \\ \frac{1-\cos(p_i B)}{1-\cos(p_i)} & \text{otherwise}. \end{cases} \qquad (75)$$

We thus arrive at

$$G(0, x') = 2B^{-4}L^{-2} \sum_{p \neq 0} \hat{p}^{-2} F(p_1) F(p_2) \left(1 - \cos(px'B)\right). \qquad (76)$$

The effective Laplacian defined in section 4 can be obtained as follows:

$$\Delta_{\text{eff}}(0, x') = l^{-2} \sum_{P} \exp(iPx') \, \tilde{\Delta}_{\text{eff}}(P). \qquad (77)$$

Here the summation over all block lattice momenta $P_i = (0, 1, ..., l-1)\frac{2\pi}{l}$, and $\tilde{\Delta}_{\text{eff}}(P)$ is given by

$$\tilde{\Delta}_{\text{eff}}(P) = \begin{cases} 0 & \text{if } P = 0, \\ \left(\frac{1}{2} \sum_{x'} \exp(-iPx') G(0, x')\right)^{-1} & \text{otherwise}. \end{cases} \qquad (78)$$

## Acknowledgements

Part of this work is contained in the thesis of M.H. [53], who would like to thank Steffen Meyer for being a solicitous advisor. M.H. also gratefully acknowledges support by the Deutsche Forschungsgemeinschaft by grant No. ME 567/3-3.

M.H. and K.P. enjoyed support by the German-Israeli Foundation for Research and Development (GIF).

All of us appreciated very much the stimulating discussions with Hans Gerd Evertz, Gidi Lana and Sorin Solomon. Two of us (M.H. and K.P.) want to thank Sorin Solomon for the warm hospitality extended to us during our visits in Israel.

Let us finally express our gratitude to the HLRZ in Jülich and to the RHRK Kaiserslautern, where most of the simulations were done.

# References

[1] W.K. Burton, N. Cabrera, and F.C. Frank, Phil. Trans. Roy. Soc. (London) 243 A, 299 (1951).

[2] J.C. Heyraud and J.J. Métois, Surf. Sci. 128, 334 (1983).

[3] I.K. Robinson, E. Vlieg, H. Hornis, and E.H. Conrad, Phys. Rev. Lett. 67, 1890 (1991).

[4] F. Gallet, S. Balibar, and E. Rolley, J. Phys. France 48, 369 (1987), and references cited therein.

[5] J.D. Weeks, G.H. Gilmer, and H.J. Leamy, Phys. Rev. Lett. 31, 549 (1973).

[6] C. Itzykson, M.E. Peskin, and J.B. Zuber, Phys. Lett. 95 B, 259 (1980).

[7] A. Hasenfratz, E. Hasenfratz, and P. Hasenfratz, Nucl. Phys. B180, 353 (1981).

[8] M. Lüscher, G. Münster, and P. Weisz, Nucl. Phys. B180, 1 (1981).

[9] G. Münster and P. Weisz, Nucl. Phys. B180, 13 (1981).

[10] G. Münster and P. Weisz, Nucl. Phys. B180, 330 (1981).

[11] M. Lüscher, Nucl. Phys. B180, 317 (1981).

[12] J.M. Drouffe and J.B. Zuber, Nucl. Phys. B180, 253 (1981).

[13] J.M. Drouffe and J.B. Zuber, Nucl. Phys. B180, 264 (1981).

[14] R. Savit, Rev. Mod. Phys. 52, 453 (1980), and references cited therein.

[15] To give only a few references:
J.M. Kosterlitz and D.J. Thouless, J. Phys. C6, 1181 (1973);
J.M. Kosterlitz, J. Phys. C7, 1046 (1974);
S.T. Chui and J.D. Weeks, Phys. Rev. B14, 4978 (1976);
J.V. José, L.P. Kadanoff, S. Kirkpatrick, and D.R. Nelson, Phys. Rev. B16, 1217 (1977);
T. Ohta and K. Kawasaki, Prog. Theor. Phys. 60, 365 (1978);
D.J. Amit, Y.Y. Goldschmidt, and G. Grinstein, J. Phys. A13, 585 (1980).

[16] D.B. Abraham, 'Surface Structures and Phase Transitions – Exact Results', in: 'Phase Transitions and Critical Phenomena', Vol. 10, C. Domb and J.L. Lebowitz, eds., Academic, London, 1986;

H. van Beijeren and I. Nolden, 'The Roughening Transition', in: 'Topics in Current Physics', Vol. 43, W. Schommers and P. van Blankenhagen, eds., Springer, Heidelberg, 1987.

[17] J.M. Fröhlich and T. Spencer, Phys. Rev. Lett. 46, 1006 (1981);

J.M. Fröhlich and T. Spencer, Commun. Math. Phys. 81, 527 (1981);

J.M. Fröhlich and T. Spencer, in: 'Scaling and Selfsimilarity in Physics', J. Fröhlich, ed., Birkhäuser, Basel, 1984.

[18] A. Patrascioiu and E. Seiler, Phys. Rev. Lett. 60, 875 (1988);

E. Seiler, I.O. Stamatescu, A. Patrascioiu, and V. Linke, Nucl. Phys. B305, 623 (1988).

[19] K. Binder, in: 'Phase Transitions and Critical Phenomena', Vol. 5b, C. Domb and M.S. Green, eds., Academic Press, London, 1976.

[20] R.H. Swendsen, Phys. Rev. B15, 5421 (1977).

[21] R.H. Swendsen, Phys. Rev. B17, 3710 (1978).

[22] W.J. Shugard, J.D. Weeks, and G.H. Gilmer, Phys. Rev. Lett. 41, 1399 (1978).

[23] R.H. Swendsen, Phys. Rev. B25, 2019 (1982).

[24] W.J. Shugard, J.D. Weeks, and G.H. Gilmer, Phys. Rev. B25, 2022 (1982).

[25] E. Bürkner and D. Stauffer, Z. Phys. B53, 241 (1983).

[26] K.K. Mon, S. Wansleben, D.P. Landau, and K. Binder, Phys. Rev. Lett. 60, 708 (1988).

[27] K.K. Mon, D.P. Landau, and D. Stauffer, Phys. Rev. B42, 545 (1990).

[28] H.G. Evertz, M. Hasenbusch, M. Marcu, K. Pinn, and S. Solomon, J. Phys. France 1, 1669 (1991).

[29] M. Hasenbusch and K. Pinn, Physica A192, 342 (1993).

[30] H.G. Evertz, M. Hasenbusch, M. Marcu, and K. Pinn, Physica A199, 31 (1993).

[31] M. Hasenbusch and K. Pinn, Physica A203, 189 (1994).

[32] S. Miyashita, H. Nishimori, A. Kuroda, and M. Suzuki, Prog. Theor. Phys. 60, 1669 (1978).

[33] S. Miyashita, Prog. Theor. Phys. 63, 797 (1980).

[34] S. Miyashita, Prog. Theor. Phys. 65, 1595 (1981).

[35] J. Tobochnik and G.V. Chester, Phys. Rev. B20, 3761 (1979).

[36] H. Betsuyaka, Physica A106, 311 (1981).

[37] J.E. Van Himbergen and S. Chakravarty, Phys. Rev. B23, 359 (1981).

[38] J.E. Van Himbergen, Phys. Rev. B25, 5977 (1982).

[39] J.E. Van Himbergen, J. Phys. C17, 5039 (1984).

[40] G.G. Batrouni, G.R. Katz, A.S. Kronfeld, G.P. Lepage, B. Svetitsky, and K.G. Wilson, Phys. Rev. D32, 2736 (1985).

[41] J. Kogut and J. Polonyi, Nucl. Phys. B265, 313 (1986).

[42] P. Harten and P. Suranyi, Nucl. Phys. B265, 615 (1986).

[43] J.F. Fernández, M.F. Ferreira, and J. Stankiewicz, Phys. Rev. B34, 292 (1986).

[44] M.S.S. Challa and D.P. Landau, Phys. Rev. B33, 437 (1986).

[45] E. Dagatto and J.B. Kogut, Phys. Rev. Lett. 58, 299 (1987).

[46] H. Weber and P. Minnhagen, Phys. Rev. B37, 5986 (1988).

[47] W. Bernreuther and M. Göckeler, Phys. Lett. B214, 109 (1988).

[48] R.G. Edwards and A.D. Sokal, Phys. Rev. D38, 2009 (1988).

[49] R. Gupta, J. Delapp, G.G. Batrouni, and G.C. Fox, Phys. Rev. Lett. 61, 1996 (1988).

[50] U. Wolff, Nucl. Phys. B322, 759 (1989).

[51] W. Janke and K. Nather, Phys. Lett. A157, 11 (1991).

[52] L. Biferale and R. Petronzio, Nucl. Phys. B328, 677 (1989).

[53] M. Hasenbusch, PhD thesis, Universität Kaiserslautern, 1992.

[54] M. Hasenbusch, M. Marcu, and K. Pinn, Nucl. Phys. B (Proc. Suppl.) 26 (1992) 598.

[55] R.J. Baxter, 'Exactly Solved Models in Statistical Mechanics', Academic, London, 1982.

[56] E.H. Lieb, Phys. Rev. 162, 162 (1967).

[57] E.H. Lieb and F.Y. Wu, <u>in</u>: 'Phase Transitions and Critical Phenomena', C. Domb and M.S. Green, eds., Vol. 1, Academic, London, 1972.

[58] R.H. Swendsen and J.S. Wang, Phys. Rev. Lett. 58, 86 (1987).

[59] D. Kandel, E. Domany, and A. Brandt, Phys. Rev. Lett. 60, 1591 (1988);

[60] U. Wolff, Phys. Rev. Lett. 62, 361 (1989).

[61] R.C. Brower and P. Tamayo, Phys. Rev. Lett. 62, 1087 (1989).

[62] H.G. Evertz, M. Hasenbusch, M. Marcu, K. Pinn, and S. Solomon, Phys. Lett. B254, 185 (1991);

H.G. Evertz, M. Hasenbusch, M. Marcu, K. Pinn, and S. Solomon, Nucl. Phys. B (Proc. Suppl.) 20, 110 (1991);

H.G. Evertz, M. Hasenbusch, M. Marcu, K. Pinn, and S. Solomon, in: 'Proceedings of the Workshop on Fermion Algorithms', Jülich 1991, H.J. Herrmann and F. Karsch, eds., World Scientific, Singapore, 1991.

[63] H.G. Evertz, M. Hasenbusch, G. Lana, M. Marcu, and K. Pinn, Phys. Rev. B46, 10472 (1992).

[64] H.G. Evertz, G. Lana, and M. Marcu, Phys. Rev. Lett. 70 875 (1993).

[65] H. van Beijeren, Phys. Rev. Lett. 38, 993 (1977).

[66] L.P. Kadanoff, Physics 2, 263 (1966).

[67] K.G. Wilson, Physica 73, 119 (1974);
K.G. Wilson and J. Kogut, Phys. Rep. C12, 75 (1974).

[68] C. Itzykson and J.M. Drouffe, 'Statistical Field Theory', Cambridge University Press, 1989.

[69] T. Niemeijer and J.M.J. Van Leeuwen, in: 'Phase transitions and Critical Phenomena', Vol. 6, C. Domb and M.S. Green, eds., Academic, London, 1976.

[70] S.K. Ma, Phys. Rev. Lett. 37, 461 (1976).

[71] S.H. Shenker and J. Tobochnik, Phys. Rev. B22, 4462 (1980).

[72] K.G. Wilson, in: 'Recent developments of gauge theories', G. 't Hooft et al., eds., Plenum, New York, 1980.

[73] M.P. Nightingale, Physica A83, 561 (1976);
K. Binder, Z. Phys. B 43, 119 (1981).

[74] R.H. Swendsen and A.M. Ferrenberg, Phys. Rev. Lett. 61, 2635 (1988).

[75] C.F. Baillie and R. Gupta, Nucl. Phys. B (Proc. Suppl.) 20, 669 (1991).